\documentclass[aps,prb,twocolumn,notitlepage,showkeys,showpacs,superscriptaddress,longbibliography,preprintnumbers]{revtex4-2}
\usepackage{epsf}
\usepackage{bm}
\usepackage{times}
\usepackage{amsfonts}
\usepackage{amssymb}
\usepackage{amsmath,mathtools}
\usepackage{array}
\usepackage{enumitem,dsfont,here,physics}
\usepackage{dcolumn,multirow}
\usepackage{tikz-cd}
\usepackage{bbding,bbold}
\usepackage[colorlinks=true,citecolor=blue,linkcolor=blue]{hyperref}
\usepackage[normalem]{ulem}
\usepackage{longtable}
\usepackage{comment}
\usepackage[whole]{bxcjkjatype}
\usepackage{braket}
\usepackage{multirow}

\graphicspath{{./fig/}}

\newcommand{\bk}{\bm{k}}

\newcommand{\mZ}{\mathbb{Z}}

\newcommand{\ii}{{\mathrm{i}}}

\graphicspath{{./fig/}}

\begin{document}
\title{Optical response of edge modes in time-reversal symmetric topological superconductors}
\author{Hirokazu Kobayashi}
\email{kobayashi-hirokazu544@g.ecc.u-tokyo.ac.jp}
\affiliation{Department of Applied Physics, University of Tokyo, Tokyo 113-8656, Japan}
\author{Han Bi}
\affiliation{International Center for Quantum Design of Functional Materials (ICQD), Hefei National Research Center for Interdisciplinary Sciences at the Microscale, University of Science and Technology of China, Hefei, Anhui 230026, China}
\author{James Jun He}
\affiliation{Hefei National Laboratory, Hefei, Anhui 230088, China}
\affiliation{International Center for Quantum Design of Functional Materials (ICQD), Hefei National Research Center for Interdisciplinary Sciences at the Microscale, University of Science and Technology of China, Hefei, Anhui 230026, China}
\author{Seishiro Ono}
\affiliation{Interdisciplinary Theoretical and Mathematical Sciences Program (iTHEMS), RIKEN, Wako 351-0198, Japan}

\preprint{RIKEN-iTHEMS-Report-24}

\begin{abstract}
    Topological superconductors and Majorana edge modes at their boundaries have been theoretically predicted.
    However, their experimental observation remains controversial.
    Recent theoretical studies suggest that chiral Majorana edge modes exhibit distinct spatially-resolved optical conductivity compared to chiral Dirac edge modes.
    In this work, we investigate the optical conductivity and spatially-resolved optical conductivity induced by Majorana edge modes and Dirac edge modes under time-reversal symmetry and crystalline symmetry.
    We conduct numerical calculations and analytical calculations with edge effective theory for two-dimensional ${\mathbb Z}_2$ topological insulators, strong topological superconductors, and topological crystalline superconductors.
    Our results show that even under time-reversal symmetry and crystalline symmetry, Majorana edge modes and Dirac edge modes exhibit different optical responses.
\end{abstract}
\maketitle

\section{Introduction}
Since the theoretical discovery of Majorana fermions in topological superconductors, their experimental realization has been awaited for the past decades.
Representative examples of theoretical predictions include edge modes of one-dimensional $p$-wave superconductors~\cite{Kitaev_Majorna_2001} and two-dimensional $p+\ii p$ superconductors~\cite{Read-Green_Majorana_2000}.
Indeed, enormous efforts have been made to realize these systems~\cite{Fu-Kane_Majorana_2D_2008, Alicea_Majorana_2D_2010, Sau-Lutchyn-Sarma_Majorana_2D_2010, Oreg-Rafael-Felix_Majorana_1D_2010, Lutchyn-Sau-Sarma_Majorana_1D_2010, Cook-Franz_Majorana_1D_2011, Choy-Beenakker_Majorana_1D_2011, Nadj-Yazdani_Majorana_1D_2013, Nakosai-Tanaka-Nagaosa_Majorana_2D_2013, Joel-Teemu_Majorana_2D_2015}.
One of the most difficult challenges is the detection of Majorana fermions.
There exist a lot of proposals to solve this problem, such as zero-bias peaks in the differential conductance, the fractional Josephson effect, and the half-integer conductance plateau~\cite{Fu-Kane_Majorana_detection_2009, Law-Lee-Ng_Majorana_detection_2009, Flensberg_Majorana_detection_2010, 
Falko-Felix_Majorana_detection_2012, Mi-Beenakker_Majorana_detection_2013, Wang-Zhang_Majorana_detection_2015}, and actual experiments have been conducted~\cite{Mourik-Kouwenhoven_Majorana_experiment_2012, Das_Majorana_experiment_2012, Deng_Majorana_experiment_2012, Rokhinson_Majorana_experiment_2012}.
However, results obtained using the existing methods are sometimes controversial due to the difficulty in distinguishing effects caused by Majorana fermions from those arising from other physical phenomena~\cite{Liu-Patrick_Majorana_difficulty_2012, Pikulin-Beenakker_Majorana_difficulty_2012, Ji-Wen_Majorana_difficulty_2018, Huang-Sau_Majorana_difficulty_2018, Moore-Tewari_Majorana_difficulty-2018, Chiu-Sarma_Majorana_difficulty_2019}.
Therefore, it is still important to develop new schemes to detect Majorana fermions.

One of the promising routes for realizing topological superconductors is the proximity effect~\cite{Fu_Kane_PRL2008,Fu_Kane_PRB2009}.
For example, topological superconductivity has been proposed in hybrid systems composed of topological insulators and superconductors~\cite{Qi-Majorana_2D_2010, Trang:2020aa}.
Another proposal is the surface superconductivity proximity-induced by bulk superconductivity of topological materials~\cite{STSC_iron, STSC_iron2, STSC_iron3}. 
In these cases, in addition to the difficulty of detecting Majorana fermions, it is crucial to distinguish between Majorana edge modes and Dirac edge modes, which originate from topological superconductors and insulators, respectively.
Recently, Ref.~\cite{James_Majorana} has theoretically proposed that the spatially-resolved optical conductivity, which measures the conductivity when light is applied to only part of the system, can distinguish between the Majorana edge mode in the $p+\ii p$ superconductor and the Dirac edge mode in a Chern insulator. 
Although both $p+\ii p$ superconductor and Chern insulator exhibit similar linear dispersions in their edge modes, the difference between Majorana edge modes and Dirac edge modes leads to distinct behaviors in the spatially-resolved optical conductivity. 

Topological superconductivity can coexist with time-reversal and/or crystalline symmetries.
In such cases, unlike in $p+\ii p$ superconductors, the number of edge states is generally greater than one, which can potentially lead to more complex optical responses.
For example, it is known that the lowest-energy optical excitation can occur in multiband superconducting systems.
Since quantum spin Hall insulators do not exhibit optical excitations between edge states at the same momentum~\cite{Han-James}, the presence of such low-energy optical excitations in bulk-gapped superconductors may serve as a signature of Majorana edge modes.
Even if such optical excitations do not occur, considering the success of spatially-resolved optical conductivity in $p+\ii p$ superconductors~\cite{James_Majorana}, it is reasonable to expect that spatially-resolved optical conductivity could also serve as a signature of Majorana edge modes in the presence of time-reversal and crystalline symmetries.
However, the optical response of Majorana edge modes in superconducting systems with time-reversal and crystalline symmetries has not been fully investigated.

To address this issue, in this work we study optical responses in various topological superconductors with time-reversal and crystalline symmetries~\cite{Periodic_table_PRB, Kitaev_bott, Ryu_2010, Shiozaki-Sato_crystalline_insulator_SC,Shiozaki-Sato-Gomi2016,Shiozaki-Sato-Gomi_mobius_twist}.
By combining symmetry analysis, numerical simulations, and analytical calculations, we compare the behavior of optical and spatially-resolved optical conductivities in systems with Majorana edge modes and Dirac edge modes.
Our strategy is as follows. 
First, we diagnose whether or not low-energy optical excitations between edge modes with the same momentum can occur based on symmetry.
If such optical excitations are allowed, the behavior is distinct from that in the two-dimensional ${\mathbb Z}_2$ topological insulator~\cite{Kane-Mele_Z2_topo_insulator, Bernevig-Hughes-Zhang_Z2_topo_insulator, Liang-Kane_Z2_topo_insulator, Hasan-Kane_topo_colloquim}, as mentioned above. 
If such optical excitations are absent, we then examine the behavior of the spatially-resolved optical conductivity.
We reveal that the spatially-resolved optical conductivity for topological superconductors generally depends on the energy of the incident photons in the low-energy regime, while that for $\mZ_2$ topological insulators does not depend. 
This finding suggests that optical conductivity and spatially-resolved optical conductivity could serve as valuable tools for the experimental observation of Majorana edge modes in a broader range of topological superconductors.

This paper is organized as follows.
In Sec.~\ref{sec:framework}, we introduce our framework utilized throughout this paper.
In Sec.~\ref{sec:results}, we present the calculation results for the optical conductivity and spatially-resolved optical conductivity for the two dimensional ${\mathbb Z}_2$ topological insulator, strong topological superconductor, and topological crystalline superconductor.
The conclusion is provided in Sec.~\ref{sec:conclusion}.

\section{Framework}
\label{sec:framework}
In this section, we introduce several fundamental quantities and techniques used in this work. 
In Sec.~\ref{sec:current}, we define the current operator in the mean-field Hamiltonian where $\text{U}(1)$ symmetry is absent.
In Sec.~\ref{sec:optical_conductivity}, we review two ways to compute the spatially-resolved optical conductivity.
In Sec.~\ref{sec:edge}, we discuss how to derive the effective edge theory from a given bulk Hamiltonian, and then we show how to compute the spatially-resolved optical conductivity based on the effective edge theory.
In Sec.~\ref{sec:q0}, we review the relation between symmetry and optical conductivity and discuss how to apply this relation to the edge mode.

\subsection{Bogoliubov-de Gennes Hamiltonian and Current operator}
\label{sec:current}
In this work, we consider translational invariant superconductors described by Bogoliubov-de Gennes (BdG) Hamiltonians in two dimensions
\begin{gather}
    \hat{H}=\frac{1}{2}\sum_{\bm k}\hat{\bm \Psi}^{\dagger}_{\bm k}H_{\bm k}\hat{\bm \Psi}^{}_{\bm k},\\
	H_{\bm k}=
    \begin{pmatrix}
	h_{\bm k} & \Delta_{\bm k}\\
	\Delta_{\bm k}^{\dagger} & -h_{-{\bm k}}^{\top}
    \end{pmatrix},
    \label{eq:BdG_Hamiltonian}
\end{gather}
where $h_{\bm k}$ and $\Delta_{\bm k}\ (\Delta_{\bk}^{\top} = -\Delta_{-\bk})$ are a normal-phase Hamiltonian and a superconducting order parameter, respectively.
Here, $\hat{\bm \Psi}^{\dagger}_{\bm k}= (\hat{\bm c}^{\dagger}_{{\bm k}}\ \hat{\bm c}^{\top}_{{-\bm k}})$ is composed of fermionic creation and annihilation operators $\hat{\bm c}^{\dagger}_{{\bm k}}$ and $\hat{\bm c}_{{\bm k}}$. 
The BdG Hamiltonian inherently possesses particle-hole symmetry $C$, satisfying 
\begin{align}
    U(C)H_{\bm k}^{\top}=-H_{-{\bm k}}U(C);\quad 
    U(C) = \begin{pmatrix}
        & \mathds{1} \\
        \mathds{1} &
    \end{pmatrix},
\end{align}
where $\mathds{1}$ is the identity matrix. 
In this work, we always consider systems with time-reversal symmetry $T$, where the Hamiltonian satisfies $U(T)H_{\bm k}^{\ast} =H_{-{\bm k}}U(T)$.

For later convenience, we introduce the real-space Hamiltonian via Fourier transformation as follows:
\begin{align}
    \label{eq:Fourier}
    \hat{\bm \Psi}^{}_{{\bm k}}&=\frac{1}{\sqrt{N}}\sum_{\bm R}e^{-\ii{\bm k}\cdot{\bm R}}\hat{\tilde{\bm \Psi}}^{}_{{\bm R}},
\end{align}
where $N=L_x L_y$ is the system size, and ${\bm R}$ is a position of a unit cell origin.
After substituting Eq.~\eqref{eq:Fourier} into Eq.~\eqref{eq:BdG_Hamiltonian}, we obtain
\begin{gather}
	\hat{H}=\frac{1}{2}\sum_{{\bm R},{\bm R}'}\hat{\tilde{\bm \Psi}}^{\dagger}_{\bm R}\tilde{H}_{{\bm R},{\bm R}'}\hat{\tilde{\bm \Psi}}^{}_{{\bm R}'},\\
	\tilde{H}_{{\bm R},{\bm R}'}=
	\begin{pmatrix}
		\tilde{h}_{{\bm R},{\bm R}'} & \tilde{\Delta}_{{\bm R},{\bm R}'}\\
		\tilde{\Delta}_{{\bm R},{\bm R}'}^{\dagger} & -\tilde{h}_{{\bm R},{\bm R}'}^*
	\end{pmatrix}
  =\sum_{\bm k}H_{\bm k}e^{\ii{\bm k}\cdot({\bm R}-{\bm R}')}.
\end{gather}

\begin{figure}[t]
	\centering
    \includegraphics[width=0.95\columnwidth]{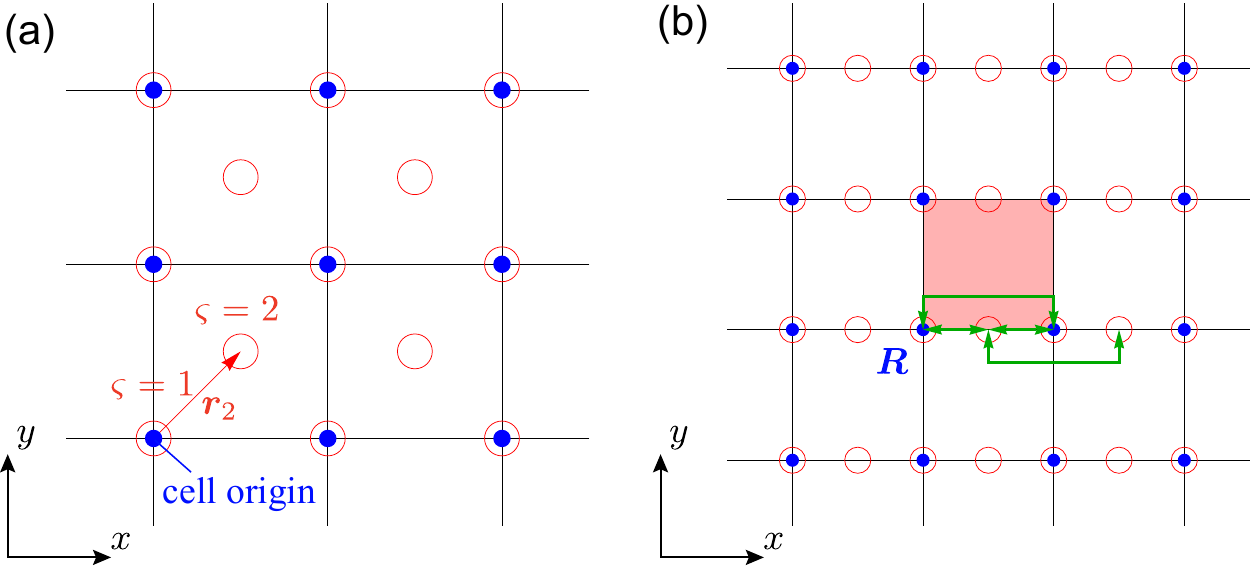}
	\caption{
        Illustration of the current in a two-dimensional lattice system.
        Each blue dot represents the origin of each unit cell, and the red circles represent the positions where orbital degrees of freedom are defined.
        (a) The orbital degrees of freedom $\varsigma$ and its coordinate ${\bm r}_{\varsigma}$ from the unit cell origin.
        (b) An example of the current along the $x$-direction.
        The links contributing to the current of the highlighted unit cell are shown with green arrows.
	}
	\label{fig:current_link}
\end{figure}
Here, we explain how to define the local current operator for the BdG Hamiltonians through $\text{U}(1)$-gauge fields.
To achieve this, we introduce $\text{U}(1)$-symmetric Hamiltonians under the gauge fields in two dimensions. 
For convenience, $\varsigma$ denotes the orbital degrees of freedom in a unit cell, and ${\bm r}_{\varsigma}$ is a coordinate of the degree of freedom $\varsigma$ from the unit cell origin [See Fig.~\ref{fig:current_link}(a) for an illustration].
The gauge field is defined on a link between two positions ${\bm R} + \bm{r}_{\varsigma}$ and ${\bm R}' + \bm{r}_{\varsigma'}$, which is represented by ${\bm A}_{{\bm R}'\varsigma',{\bm R}\varsigma}={\bm A}_{{\bm R}\varsigma,{\bm R}'\varsigma'}$.
The hopping under the gauge field is then expressed as
\begin{align}
  \left[\tilde{h}_{{\bm R},{\bm R}'}({\bm A})\right]_{\varsigma,\varsigma'}=\left[\tilde{h}_{{\bm R},{\bm R}'}\right]_{\varsigma,\varsigma'}e^{\ii {\bm A}_{{\bm R}\varsigma,{\bm R}'\varsigma'}\cdot ({\bm R}+{\bm r}_{\varsigma}-{\bm R}'-{\bm r}_{\varsigma'})}.
\end{align}
Similar to Refs.~\cite{Yamamoto_current_definition,Sven-Martin_current_definition,Watanabe_Bloch}, we define the local current operator for $\text{U}(1)$-symmetric Hamiltonians by the derivative of the gauge field.

In contrast, the BdG Hamiltonian does not possess $\text{U}(1)$ symmetry.
Nonetheless, considering the gauge field couples only to the normal phase, we define the current operator on the link for BdG Hamiltonians by the derivative of the gauge field~\cite{Ahn-Nagaosa_NC,Furusaki_Chiral_HE,Papaj_SC_current}:
\begin{align}
  \label{eq:local_current_def}
	\hat{\tilde{j}}^i_{{\bm R}\varsigma,{\bm R}'\varsigma'}\coloneqq -\left.\frac{\partial \hat{H}({\bm A})}{\partial A^i_{{\bm R}\varsigma,{\bm R}'\varsigma'}}\right|_{{\bm A}={\bm 0}},
\end{align}
where
\begin{gather}
	\hat{H}({\bm A})=\frac{1}{2}\sum_{{\bm R},{\bm R}'}\hat{\tilde{\bm \Psi}}^{\dagger}_{\bm R}\begin{pmatrix}
		\tilde{h}_{{\bm R},{\bm R}'}({\bm A}) & \tilde{\Delta}_{{\bm R},{\bm R}'}\\
		\tilde{\Delta}_{{\bm R},{\bm R}'}^{\dagger} & -[\tilde{h}_{{\bm R},{\bm R}'}({\bm A})]^*
	\end{pmatrix}\hat{\tilde{\bm \Psi}}^{}_{{\bm R}'}.
\end{gather}
Furthermore, the local current operator at unit cell ${\bm R}$ is defined as the sum of the link current operators:
\begin{align}
  \label{eq:local_current}
  \hat{\tilde{j}}^i_{\bm R}\coloneqq-\hspace{-0.5cm}\sum_{\substack{{\bm R}',\varsigma,\varsigma'\\(({\bm R}'+\bm{r}_{\varsigma'})_i > ({\bm R} +\bm{r}_{\varsigma })_i)}}\hspace{-0.5cm}\hat{\tilde{j}}^i_{{\bm R}\varsigma,{\bm R}'\varsigma'}.
\end{align}
For instance, let us consider the current along $x$-direction for a model with nearest- and second-nearest-neighbor hoppings in a square lattice, as illustrated in Fig.~\ref{fig:current_link}(b).
In this case, the links that contribute to the local current at unit cell ${\bm R}$ are those connecting the orbitals within unit cell ${\bm R}$, and the links between orbitals in unit cells ${\bm R}$ and ${\bm R}+{\bm a}_x$:
\begin{align}
  \hat{\tilde{j}}_{\bm R}^x&=\hat{\bm \Psi}^{\dagger}_{\bm R}\tilde{\iota}^x_{{\bm R},{\bm R}+{\bm a}_x}\hat{\bm \Psi}^{}_{{\bm R}+{\bm a}_x}
  +\hat{\bm \Psi}^{\dagger}_{{\bm R}+{\bm a}_x}\tilde{\iota}^x_{{\bm R}+{\bm a}_x,{\bm R}}\hat{\bm \Psi}^{}_{{\bm R}}
  \notag\\
  &\qquad +\hat{\bm \Psi}^{\dagger}_{\bm R}\tilde{\iota}^x_{{\bm R},{\bm R}}\hat{\bm \Psi}^{}_{{\bm R}},\\
	\left[\tilde{\iota}^x_{{\bm R},{\bm R}'}\right]_{\varsigma,\varsigma'}&=-\ii ({\bm R}+{\bm r}_{\varsigma}-{\bm R}'-{\bm r}_{\varsigma'})_x
	\begin{pmatrix}
		\tilde{h}_{{\bm R},{\bm R}'} & 0\\
		0 & \tilde{h}_{{\bm R},{\bm R}'}^*
	\end{pmatrix}_{\varsigma,\varsigma'},
\end{align}
where ${\bm a}_i$ is the primitive lattice vector along the $i$-direction.

\subsection{Spatially-resolved optical conductivity}
\label{sec:optical_conductivity}
\begin{figure}[t]
	\centering
	\includegraphics[width=0.95\columnwidth]{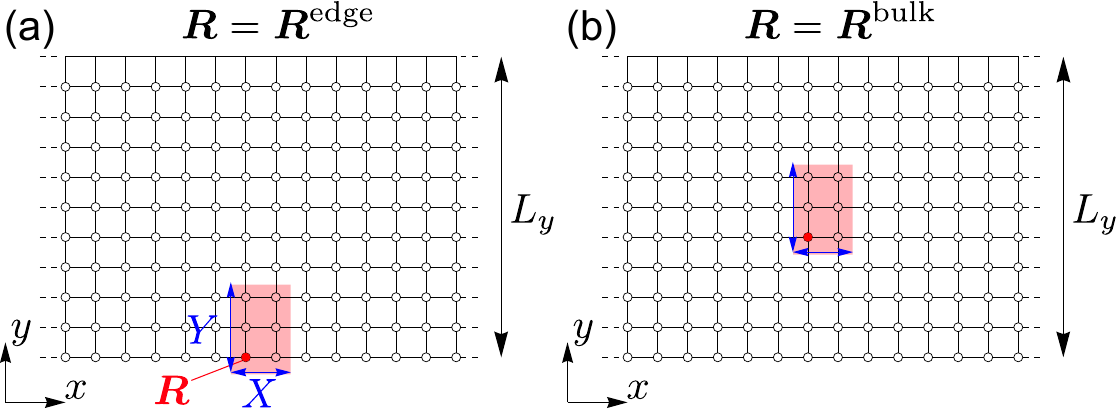}
	\caption{
        Illustration of systems for spatially-resolved optical conductivity.
        Each white circle represents the origin of each unit cell.
        We consider systems defined on $[0, L_x) \times [0, L_y)$, where $L_x$ is sufficiently large while $L_y$ is finite as mentioned below. 
        We impose periodic and open boundary conditions in the $x$ direction and the $y$ direction, respectively.
        Light is applied to the shaded region specified by ${\bm R}=(R_x, R_y)$, and the size of the region can be adjusted by $(X, Y)$.
        We analyze the spatially-resolved optical conductivity $\tilde{\sigma}_{xx}(\omega,{\bm R})$ in this region.
        In our numerical simulations, we set $L_y=30$, $X = 1$, and $Y = 5$.
        (a) Light is applied to the edge with ${\bm R}={\bm R}^{\rm edge}=(R_x^{\rm edge}, R_y^{\rm edge})$, where $R_y^{\rm edge}=0$ and $R_x^{\rm edge}$ is arbitrarily due to the presence of translational symmetry.
        (b) Light is applied to the bulk with ${\bm R}={\bm R}^{\rm bulk}=(R_x^{\rm bulk}, R_y^{\rm bulk})$, where $R_y^{\rm bulk}=\lceil (L_y-Y)/2 \rceil$ and $R_x^{\rm bulk}$ is also arbitrarily.
	}
	\label{fig:local_optical_conductivity}
\end{figure}

In the following, we discuss the spatially-resolved optical conductivity, where light is applied to only part of a system as illustrated in Fig.~\ref{fig:local_optical_conductivity}.
When light is applied starting from ${\bm R}=(R_x,R_y)$ over the region $[R_x,R_x+X-1]\times[R_y,R_y+Y-1]$, the spatially-resolved optical conductivity is defined within the framework of linear response theory~\cite{Kubo_linear_response,Mahan-Many_particle} as
\begin{align}
  \tilde{\sigma}_{ij}(\omega,{\bm R})\hspace{-0.05cm}&\coloneqq \frac{1}{2\pi\omega}\int_{0}^{\infty}\hspace{-0.18cm}dt\ e^{\ii\omega t}\langle[e^{\ii\hat{H}t}\hat{\tilde{J}}^i({\bm R})e^{-\ii\hat{H}t},\hat{\tilde{J}}^j({\bm R})]\rangle,
\end{align}
where $\hat{\tilde{J}}^i({\bm R})$ is the current operator in the $i$-direction within the illuminated region and is expressed as the sum of the local current operators defined in Eq.~\eqref{eq:local_current}:
\begin{gather}
    \hat{\tilde{J}}^i({\bm R})=\frac{1}{X}\sum_{X'=0}^{X-1}\sum_{Y'=0}^{Y-1}\hat{\tilde{j}}^i_{{\bm R}+X'{\bm a}_x+Y'{\bm a}_y}.
\end{gather}

Hereafter, we focus on $\tilde{\sigma}_{xx}(\omega,{\bm R})$ in systems under the periodic boundary condition along the $x$-direction with a sufficiently large system size $L_x$ and the open boundary condition along the $y$-direction with a finite system size $L_y$.

The real part of the spatially-resolved optical conductivity can be computed by Green's function~\cite{James_Majorana, Mahan-Many_particle, Rammer-Smith_sigma_Green}:
\begin{align}
    \label{eq:sigma_Green}
    {\rm Re}[\tilde{\sigma}_{xx}(\omega,{\bm R})]&=\frac{1}{2\pi^2\omega}\int_{-\infty}^{\infty}d\varepsilon\left(f(\varepsilon)-f(\varepsilon+\omega)\right)\notag\\
	&\quad \times{\rm Tr}\left[\hat{A}(\varepsilon+\omega)\hat{\tilde{J}}^x({\bm R})\hat{A}(\varepsilon)\hat{\tilde{J}}^x({\bm R})\right],
\end{align}
where $f(\varepsilon)$ denotes the Fermi-Dirac distribution, and $\hat{A}(\varepsilon)$ represents the spectral function
\begin{align}
  \hat{A}(\varepsilon)&\coloneqq -\frac{1}{2\ii}\left[\hat{G}^{r}(\varepsilon)-\hat{G}^{a}(\varepsilon)\right].
\end{align}
For numerical calculations, we utilize the recursive Green's function method to calculate the Green's function for a given system~\cite{MacKinnon_RGF, Krsti-Zhang-Butler_RGF, Zhang-Krstic-Butler_RGF, Li-Lu_RGF, Lewenkopf-Caio-Mucciolo-Eduardo_RGF, Zhang-Liu_RGF_AIP}.
For translational invariant systems, there exists an efficient algorithm based on Schur decomposition~\cite{Wimmer_RGF}.
We also use this technique in our numerical calculations. 

Finally, we introduce the momentum-dependent optical conductivity:
\begin{align}
    \label{eq:sigma_q}
  &\quad \sigma_{xx}(\omega,q,R_y)\notag\\
  &\coloneqq \frac{1}{\omega L_x}\int_{0}^{\infty}dt\ e^{\ii\omega t}\langle [e^{\ii\hat{H}t}(\hat{j}^x_{q,R_y})^{\dagger}e^{-\ii\hat{H}t},\hat{j}^{x}_{q,R_y}]\rangle,
\end{align}
where $\hat{j}^{x}_{q,R_y}$ is the current operator in wavenumber space obtained by Fourier transform in the $x$ direction:
\begin{gather}
    \label{eq:current_q}
    \hat{j}^x_{q,R_y}=\sum_{X'}\sum_{Y'=R_y}^{R_y+Y-1}e^{\ii qX'}\hat{\tilde{j}}^x_{X'{\bm a}_x+Y'{\bm a}_y}.
\end{gather}

As shown in Appendix~\ref{app:local_momentum_opt_cond}, the momentum-dependent optical conductivity is related to the spatially-resolved optical conductivity as
\begin{align}
  \label{eq:sigma_Fourier}
  &\quad \tilde{\sigma}_{xx}(\omega,{\bm R})\notag\\
  &=\frac{1}{2\pi X^2}\sum_{X',X''=0}^{X-1}\int_{-\pi}^{\pi} \frac{dq}{2\pi}~ e^{\ii q\cdot(X'-X'')}\sigma_{xx}(\omega,q,R_y).
\end{align}
This relation enables us to compute $\tilde{\sigma}_{xx}(\omega,{\bm R})$ from $\sigma_{xx}(\omega,q,R_y)$.
As discussed in Sec.~\ref{sec:results}, it is sometimes possible to analytically compute $\sigma_{xx}(\omega,q,R_y)$ in the edge using the edge theory.

\begin{figure}[t]
	\centering
	\includegraphics[width=0.95\columnwidth]{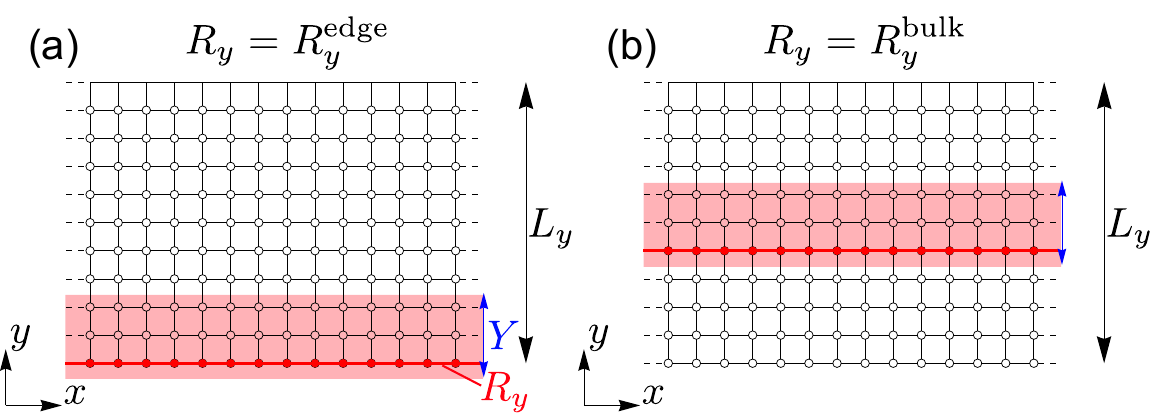}
	\caption{
        Illustration of systems for optical conductivity $\tilde{\sigma}_{xx}(\omega,q=0,R_y)$.
        The boundary conditions and system size are the same as in Fig.~\ref{fig:local_optical_conductivity}.
        In this case, light is applied uniformly across the entire $x$ direction, with a finite width $Y$ along the $y$ direction.
        In other words, the light applied region is $[0,L_x) \times [R_y,R_y+Y-1]$.
        (a) Light is applied to the edge with $R_y=R_y^{\rm edge}$.
        (b) The figure when light is applied to the bulk with $R_y=R_y^{\rm bulk}$.
	}
	\label{fig:optical_conductivity}
\end{figure}
Additionally, we can obtain the optical conductivity as the momentum-dependent optical conductivity for $q=0$.

The optical conductivity corresponds to the response when uniform light is applied in the $x$-direction, as shown in Fig.~\ref{fig:optical_conductivity}.
starting from expression \eqref{eq:sigma_q}, we can express optical conductivity as
\begin{align}
  \label{eq:sigma_band}
  \tilde{\sigma}_{xx}(\omega,q=0,R_y)&=\frac{1}{\omega}\sum_{n,m}\int_{-\pi}^{\pi} dk_x\{f(\epsilon_{nk_x})-f(\epsilon_{mk_x})\}\notag\\
  &\quad \times\left|[j_{k_x,R_y}^{x}]_{nm}\right|^2\delta(\omega-\epsilon_{mk_x}+\epsilon_{nk_x}),\notag\\
  [j_{k_x,R_y}^{x}]_{nm}&=\braket{nk_x|\hat{j}^x_{q=0,R_y}|mk_x},
\end{align}
where $\hat{H}\ket{nk_x} = \epsilon_{nk_x}\ket{nk_x}$ are the eigenstates and their corresponding energies under $x$-periodic and $y$-open boundary conditions.
This expression allows us to compute $\sigma_{xx}(\omega, q = 0,R_y)$ by diagonalizing the Hamiltonian.

\subsection{Edge theory}
\label{sec:edge}
To understand the behavior of the optical conductivity and spatially-resolved optical conductivity originating from gapless edge modes, we discuss an analytical method for calculating the momentum-dependent optical conductivity using the effective edge theory~\cite{Shen_Edge_Theory}.
For a general Hamiltonian, it is often difficult to analytically obtain the effective edge theory.
Nonetheless, when the Hamiltonian is separated in the $x$- and $y$-directions as $H_{\bm k} = h^x(k_x) + h^y(k_y)$, we can derive the effective edge theory analytically as follows.
Here, we assume that all ${\bm k}$-independent terms are included in $h^y(k_y)$.

As a preparation to obtain the effective edge theory, we perform a Fourier transform and take the continuous limit in the $y$ direction:
\begin{align}
  \hat{\bm \Psi}_{\bm k}=\frac{1}{\sqrt{L_y}}\int_{0}^{L_y} dy~ e^{-\ii k_y y}\hat{\bm \varPsi}_{k_x,y}.
\end{align}
In this basis, the BdG Hamiltonian~\eqref{eq:BdG_Hamiltonian} becomes
\begin{align}
  \hat{H}=\sum_{k_x}\int_{0}^{L_y}dy~ \hat{\bm \varPsi}_{k_x,y}^{\dagger}\left(h^x(k_x)+h^y(-\ii \partial_y)\right)\hat{\bm \varPsi}_{k_x,y}^{}.
\end{align}
Assume the system is topological and has edge modes at the boundary in the $y$ direction.
The basis for these edge modes can be obtained as the zero-energy state of the Hamiltonian in the $y$ direction $h^y(-\ii\partial_y)$.
Here, suppose that the zero-energy eigenfunction $\phi_i(y)~(i=1,\cdots,l)$ of $h^y(-\ii \partial_y)$ has the form
\begin{align}
  \phi_i(y)\propto \varphi_i(y){\bm \chi}_i,
\end{align}
where $\varphi_i(y)$ is a decaying factor and ${\bm \chi}_i$ is a normalized vector corresponding to degrees of freedom.
Here, $l$ represents the number of the edge modes.
For $\phi_i(y)$ to be the zero-energy state of $h^y(-\ii \partial_y)$, it must satisfy
\begin{align}
\label{eq:y_zero_mode}
  h^y(-\ii\partial_y)\varphi_i(y){\bm \chi}_i={\bm 0}.
\end{align}
We obtain $\varphi_i(y)$ and ${\bm \chi}_i$ by solving this equation.
Since we consider a topologically nontrivial system, $\varphi_i(y)$ decays rapidly from either the edge at $y=0$ or $y=L_y$.
We rearrange these functions such that $\phi_i~(i=1,\cdots,l')$ decay from the edge of interest, and $\phi_i~(i=l'+1,\cdots,l)$ decay from the opposite edge.
Once such pairs of $\varphi_i(y)$ and ${\bm \chi}_i$ are obtained, the effective edge theory is given by
\begin{align}
  \hat{H}^{\rm edge}=\frac{1}{2}\sum_{k_x}\hat{\bm \gamma}^{\dagger}_{k_x}H^{\rm edge}_{k_x}\hat{\bm \gamma}^{}_{k_x}.
\end{align}
The Hamiltonian matrix and basis are given by setting $\chi=({\bm \chi}_1,\cdots,{\bm \chi}_{l'})$ as follows:
\begin{align}
  H^{\rm edge}_{k_x}&=\chi^{\dagger}h^x(k_x)\chi,\\
  \hat{\gamma}^{}_{i,k_x}&=\int_{0}^{L_y}dy~ \varphi_i(y){\bm \chi}_i^{\dagger}\hat{\bm \varPsi}^{}_{k_x,y}.
\end{align}

Next, we consider the edge current.
Taking the continuous limit in the $y$ direction in~\eqref{eq:current_q} yields
\begin{gather}
  \label{eq:current_q_2}
  \hat{j}^{x}_{q,R_y}=\frac{1}{2}\sum_{k_x}\int_{R_y}^{R_y+Y}dy~\hat{\bm \varPsi}^{\dagger}_{k_x+q,y}j_{k_x,q}^{x}\hat{\bm \varPsi}^{}_{k_x,y}.
\end{gather}
In this expression, we assumed that the matrix element $j_{k_x,q}^{x}$ does not depend on $y$ due to the assumption that the Hamiltonian can be completely separated in the $x$ and $y$ directions.
By performing a projection similar to the Hamiltonian in \eqref{eq:current_q_2}, we obtain the edge current
\begin{gather}
  \hat{j}^{\rm edge}_{q}=\frac{1}{2}\sum_{k_x}\hat{\bm{\gamma}}^{\dagger}_{k_x+q}j^{\rm edge}_{k_x,q}\hat{\bm{\gamma}}^{}_{k_x},\\
  \label{eq:edge_current_q}
  j^{\rm edge}_{k_x,q}=\chi^{\dagger}j^{x}_{k_x,q}\chi.
\end{gather}

Finally, we derive the expression for the edge momentnum dependent optical conductivity using the effective edge theory.
We diagonalize the Hamiltonian $H^{\rm edge}_{k_x}$, and let $u_{nk_x}$ be an eigenvector of $H^{\rm edge}_{k_x}$ with eigenenergy $\epsilon_{nk_x}$, which satisfies $H_{k_x}^{\rm edge}u_{nk_x}=\epsilon_{nk_x}u_{nk_x}$.
When we define a fermionic operator $\hat{\psi}^{}_{nk_x}$ by $\hat{\psi}^{}_{nk_x}=u_{nk_x}^{\dagger}\hat{\bm{\gamma}}^{}_{k_x}$, the Hamiltonian and the current operator can be expressed as
\begin{gather}
  \label{eq:Heff}
  \hat{H}^{\rm edge}=\frac{1}{2}\sum_{n}\sum_{k_x}\epsilon_{nk_x}\hat{\psi}^{\dagger}_{nk_x}\hat{\psi}^{}_{nk_x},\\
  \hat{j}^{\rm edge}_{q}=\frac{1}{2}\sum_{n,m}\sum_{k_x}\hat{\psi}^{\dagger}_{nk_x+q}[j^{\psi}_{k_x,q}]_{nm}\hat{\psi}^{}_{mk_x},
\end{gather}
where $[j^{\psi}_{k_x,q}]_{nm}$ is given by $[j^{\psi}_{k_x,q}]_{nm}=u_{nk_x+q}^{\dagger}j^{\rm edge}_{k_x,q}u_{mk_x}$.
Using these matrix elements $j^{\psi}_{k_x,q}$ and the energy eigenvalues $\epsilon_{nk_x}$, the momentum-dependent optical conductivity for the edge can be obtained as
\begin{align}
    \label{eq:sigma_momentum_band}
  \sigma^{\rm edge}(\omega,q)&=\frac{1}{4\omega}\sum_{n,m}\int_{-\pi}^{\pi} dk_x\{f(\epsilon_{nk_x})-f(\epsilon_{mk_x+q})\}\notag\\
  &\quad \times\left|[j_{k_x,q}^{\psi}]_{nm}\right|^2\delta(\omega-\epsilon_{mk_x+q}+\epsilon_{nk_x}).
\end{align}
In the low-energy region, since there is no contribution from the bulk, we have $\sigma_{xx}(\omega,q,R_y)\simeq \sigma^{\rm edge}(\omega,q)$ in the edge, allowing us to investigate the behavior of the optical conductivity and spatially-resolved optical conductivity using~\eqref{eq:sigma_Fourier}.

\subsection{Symmetry and optical conductivity}
\label{sec:q0}

\begin{table}[t]
  \centering
  \caption{
    The relationship between the EAZ class and optical conductivity.
    Adapted from Ref.~\cite{Ahn-Nagaosa_NC}.
    The eigenvalue sectors of the system are classified into EAZ classes by ${\mathcal T}$, ${\mathcal C}$ and $S$.
    The columns for ${\mathcal T}^2$, ${\mathcal C}^2$ and $S^2$ indicate whether each symmetry is present within an eigenvalue sector.
    If the system lacks $P$, which results in the absence of ${\mathcal T}$ and ${\mathcal C}$, or if the symmetry transforms the state to a different eigenvalue sector, it is indicated by $0$.
    On the other hand, if each eigenvalue sector possesses ${\mathcal T}$, ${\mathcal C}$, and $S$, the projective factors $\zeta_{a}=\pm 1~(a={\mathcal C}, {\mathcal T}, S)$ are listed.
    Whether excitations are allowed for each EAZ class is given in the column titled ``Lowest energy optical excitation".
    Among those for which excitations are allowed, those marked with an asterisk require multi-band systems.
    However, the energy of the allowed optical excitation in this case may differ from the superconducting gap.
    }
	\label{table:EAZ_class}
  \begin{tabular}{ccccc}
  \hline
  EAZ class & $\mathcal{T}^2$ & $\mathcal{C}^2$ & $S^2$ & Lowest energy optical excitation\\
  \hline
  A   & 0  & 0  & 0  & Yes*\\
  AI  & 1  & 0  & 0  & Yes*\\
  AII & $-1$ & 0  & 0  & Yes*\\
  AIII & 0 & 0 & 1 & Yes\\
  D   & 0  & 1  & 1  & Yes\\
  BDI & 1  & 1  & 1  & Yes\\
  C   & 0  & $-1$ & 1  & No\\
  CI  & 1  & $-1$ & 1  & No\\
  DIII & $-1$ & 1  & 1  & Yes\\
  CII & $-1$ & $-1$ & 1  & Yes\\
  \hline
  \end{tabular}
\end{table}

We have discussed spatially-resolved optical conductivity; however, in practical experimental setups, optical conductivity tends to dominate, making it difficult to measure spatially-resolved optical conductivity when optical conductivity is present.
Therefore, it is important to determine whether the system possesses optical conductivity.
In Ref.~\cite{Ahn-Nagaosa_NC}, the authors propose a symmetry-based argument to determine the presence or absence of optical conductivity originating from the lowest energy optical excitation in bulk.
Here, we show that this argument can also be applied to the edge modes.

Before reviewing the relationship between lowest energy optical excitation and symmetry in the bulk, let us first explain a symmetry setting we consider here. 
As mentioned in Sec.~\ref{sec:current}, we always consider BdG Hamiltonians with particle-hole symmetry $C$, time-reversal symmetry $T$, and chiral symmetry $S=CT$.
When a Hamiltonian is symmetric under a crystalline symmetry group $G$, it satisfies 
\begin{align}
  U_{\bm k}(g)H_{\bm k}=H_{g{\bm k}}U_{\bm k}(g) \ \ (g\in G),
\end{align}
where $U_{{\bm k}}(g)$ is the unitary representation of $g$.
We define a unitary subgroup of $G$ by
\begin{align}
    G_{{\bm k}} = \{g \in G\ \vert\  g{\bm k} \equiv {\bm k} \},
\end{align}
where ``$\equiv$'' means $g{\bm k}={\bm k}$ up to reciprocal lattice vectors.
For simplicity, we assume that all elements in $G_{{\bm k}}$ can be simultaneously diagonalizable~\footnote{We usually consider ${\bm k}$ as generic momentum in $d$-dimension $(d \leq 3)$. In this case, this assumption is always true. }.
Then, an eigenvector ${\bm \psi}_{n{\bm k}}$ of Hamiltonian matrix $H_{\bm k}$ is also an eigenvector of $U_{\bm k}(g)\ (g \in G_{{\bm k}})$ as
\begin{align}
    U_{\bm k}(g){\bm \psi}_{n{\bm k}} = \xi_{{\bm k}}(g){\bm \psi}_{n{\bm k}},
\end{align}
where $\xi_{{\bm k}}(g)$ is an eigenvalue of $g$. 
Suppose that there exists a symmetry $P$ such that $P$ transforms momentum ${\bm k}$ into $-{\bm k}$.
In such a case, the products ${\mathcal T}\coloneqq PT$ and ${\mathcal C}\coloneqq PC$ are also symmetries and do not change momentum ${\bm k}$, whose representations satisfy
\begin{align}
    \label{eq:T2}
    &U_{{\bm k}}({\mathcal T})U^{*}_{{\bm k}}({\mathcal T}) = \zeta_{{\mathcal T}}\mathds{1}\quad (\zeta_{{\mathcal T}} = \pm 1);\\
    \label{eq:C2}
    &U_{{\bm k}}({\mathcal C})U^{*}_{{\bm k}}({\mathcal C}) = \zeta_{{\mathcal C}}\mathds{1}\quad (\zeta_{{\mathcal C}} = \pm 1).
\end{align}
Importantly, $U_{{\bm k}}({\mathcal T}){\bm \psi}_{n{\bm k}}^*$ or $U_{{\bm k}}({\mathcal C}){\bm \psi}_{n{\bm k}}^*$ can have different eigenvalues of $g$.
Such transformation properties of ${\mathcal T}$, ${\mathcal C}$, and $S$ are classified by effective Altland-Zirnbauer (EAZ) symmetry classes~\cite{Altland_AZclass, Bzdu_EAZclass} [see Table~\ref{table:EAZ_class}].

Then, we move on to the relation between lowest energy optical excitation and EAZ classes.
Optical excitations do not occur between states with different eigenvalues due to selection rules.
Thus, we focus on the optical excitation between two states with the same eigenvalue. 
The nontrivial constraints arise only when $U_{{\bm k}}({\mathcal C}){\bm \psi}_{n{\bm k}}^*$ has the same eigenvalue $\xi_{{\bm k}}(g)\ (g \in G_{{\bm k}})$ as that of ${\bm \psi}_{n{\bm k}}$.
Considering the matrix element of the current operator between the states ${\bm \psi}_{n{\bm k}}$ and $U_{\bm k}({\mathcal C}){\bm \psi}^*_{n{\bm k}}$, we obtain
\begin{align}
  {\bm \psi}_{n{\bm k}}^{\dagger}j^{i}_{\bm k}(U_{{\bm k}}({\mathcal C}){\bm \psi}_{n{\bm k}}^*)=\zeta_{{\mathcal C}}{\bm \psi}_{n{\bm k}}^{\dagger}j^{i}_{\bm k}(U_{{\bm k}}({\mathcal C}){\bm \psi}_{n{\bm k}}^*).
\end{align}
Therefore, when $\zeta_{{\mathcal C}}=-1$, no optical excitation occurs between ${\bm \psi}_{n{\bm k}}$ and $U_{{\bm k}}({\mathcal C}){\bm \psi}_{n{\bm k}}^*$, resulting in $\sigma(\omega,{\bm q}={\bm 0})=0$ in the lowest excitation energy region.
However, for class CII, i.e., $(\zeta_{{\mathcal C}}, \zeta_{{\mathcal T}})=(-1, -1)$, the lowest energy optical excitation can happen. 
This is because while optical excitation between ${\bm \psi}_{n{\bm k}}\ [U_{{\bm k}}({\mathcal T}){\bm \psi}_{n{\bm k}}^*]$ and $U_{{\bm k}}({\mathcal C}){\bm \psi}_{n{\bm k}}^*$ $[U(T)U^*(C){\bm \psi}_{n{\bm k}}]$ is prohibited, optical excitation between ${\bm \psi}_{n{\bm k}}$ and $U(T)U^*(C){\bm \psi}_{n{\bm k}}$ is not forbidden.
The results of these discussions are summarized in Table~\ref{table:EAZ_class}.

Next, we apply the above argument to the edge modes.
To do this, we need to identify the symmetries that the edge possesses.
Some symmetries in $G$, such as inversion symmetry $P$ and translation $T_y$ along the $y$ direction, do not preserve the edge.
Such symmetries cannot be considered symmetries on edge.
While such symmetries by themselves do not preserve the edge, the combinations of them do.
For example, $T_y$ and $P$ do not leave any point on the edge, but the combination $(T_y)^{L_y}P$ preserves the edge as a whole.
However, we do not consider such symmetries as those intrinsic to the edge.
To focus on symmetries intrinsic to the edge, we consider the group $G$ without translations along $y$-direction, namely, $G/\ev{T_y}$, where $\ev{T_y}\coloneqq  \{T_{y}^{n}: n\in \mathbb{Z}\}$ is a group generated only by $T_y$. 
Then, we define the subgroup $G^{\rm edge}$, which consists of the symmetries that intrinsically preserve the edge.
Their bulk representations do not depend on $k_y$, and their edge-projected representations $U^{\rm edge}_{k_x}(g)$ can be obtained similarly to Eq.~\eqref{eq:Heff} as
\begin{align}
  U^{\rm edge}_{k_x}(g)=\chi^{\dagger}U_{k_x}(g)\chi\quad (g\in G^{\rm edge}).
\end{align}
As discussed in Appendix~\ref{app:edge_symmetry}, this projected symmetry satisfies
\begin{align}
	U_{k_x}^{\rm edge}(g)H_{k_x}^{\rm edge}\left(U_{k_x}^{\rm edge}(g)\right)^{\dagger}=H_{gk_x}^{\rm edge}.
\end{align}
Similarly, the current can also be obtained as
\begin{align}
  j^{\rm edge}_{k_x}=\chi^{\dagger}j^x_{k_x}\chi,
\end{align}
which is equivalent to the expression in~\eqref{eq:edge_current_q} with $q=0$.
Considering the matrix element of $j_{k_x}^{\rm edge}$ between the edge states ${\bm \psi}_{nk_x}$ and $U^{\rm edge}_{k_x}({\mathcal C}){\bm \psi}^*_{nk_x}$, we obtain, as discussed in Appendix~\ref{app:symmetry_excitation},
\begin{align}
  &\quad {\bm \psi}_{nk_x}^{\dagger}j_{k_x}^{\rm edge}(U^{\rm edge}_{k_x}({\mathcal C}){\bm \psi}_{nk_x}^*) \notag\\
  &=\zeta_{{\mathcal C}}{\bm \psi}_{nk_x}^{\dagger}j_{k_x}^{\rm edge}(U^{\rm edge}_{k_x}({\mathcal C}){\bm \psi}_{nk_x}^*).
\end{align}
Thus, by considering the symmetries that preserve the edge for a given symmetry group of the bulk, we determine whether the optical conductivity is allowed for the edge mode.

\section{Results}
\label{sec:results}
In this section, we present the numerical results for the optical conductivity and spatially-resolved optical conductivity of strong topological superconductors and topological crystalline superconductors, both of which possess time-reversal symmetry.
We also provide analytical results derived from effective theories.
The details of the calculations are provided in Supplemental Material.

\subsection{Two-dimensional ${\mathbb Z}_2$ topological insulator}
\label{sec:qsh}
\begin{figure}[t]
	\centering
	\includegraphics[width=0.95\columnwidth]{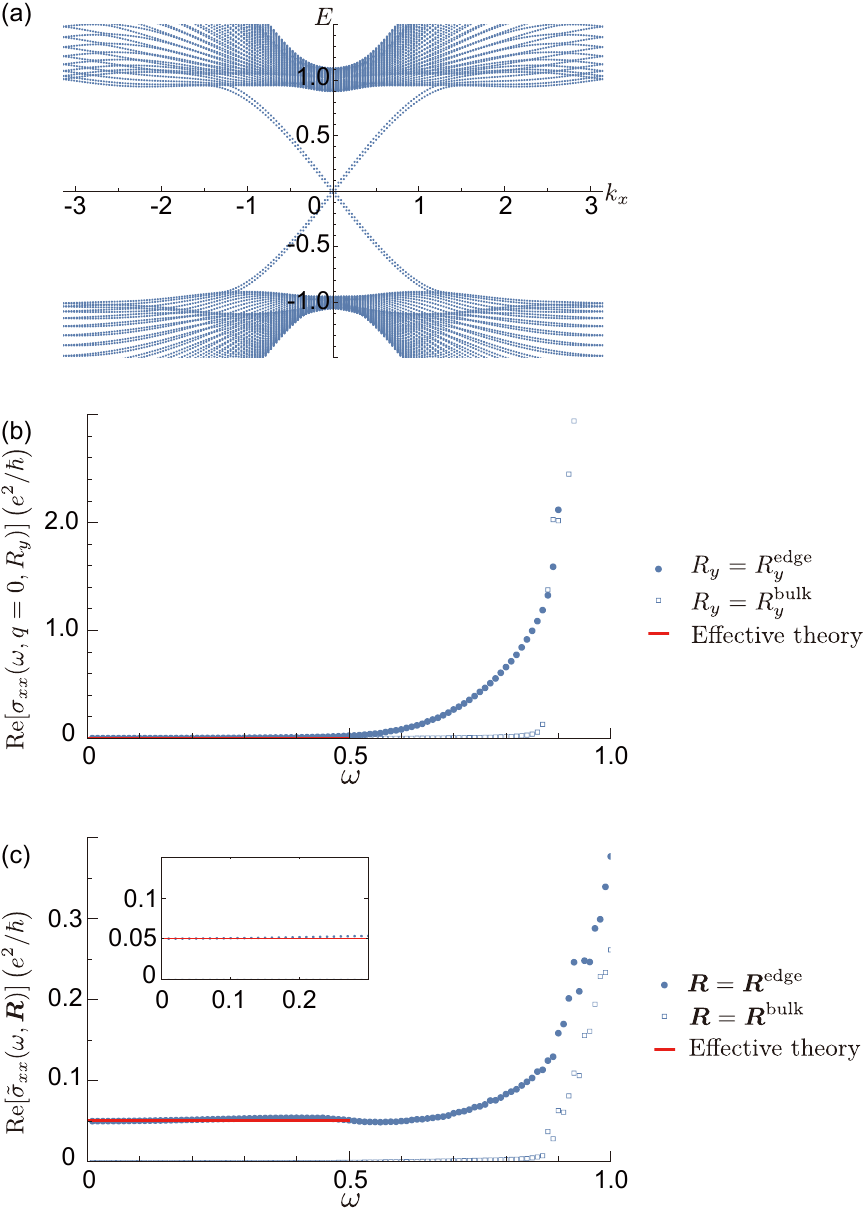}
	\caption{ 
		(a) The band structure of the two-dimensional ${\mathbb Z}_2$ topological insulator with random perturbations. The parameters used are: $t=\mu=a=1$. 
        (b) Results of numerical calculation for the optical conductivity based on Eq.~\eqref{eq:sigma_band} at $T=10^{-3}$.
        (See Supplemental Material for the explicit form of the perturbation.)
        The choice of $R_y$ is the same as in Fig.~\ref{fig:optical_conductivity}.
        (c) Results of numerical calculation for the spatially-resolved optical conductivity with $X=1$, based on Eq.~\eqref{eq:sigma_Green}  at $T=10^{-3}$ and Eq.~\eqref{eq:sigma_qsh} at $T=0$.
        The choice of ${\bm R}$ is the same as in Fig.~\ref{fig:local_optical_conductivity}.
}
	\label{fig:qsh}
\end{figure}
We discuss the optical conductivity and spatially-resolved optical conductivity of a two-dimensional ${\mathbb Z}_2$ topological insulator protected by time-reversal symmetry~\cite{Kane-Mele_Z2_topo_insulator,Bernevig-Hughes-Zhang_Z2_topo_insulator,Liang-Kane_Z2_topo_insulator,Hasan-Kane_topo_colloquim}, comparing it with topological superconductors.
The Hamiltonian of the ${\mathbb Z}_2$ topological insulator is
\begin{align}
  \hat{H}=\sum_{\bm k}\hat{\bm c}^{\dagger}_{\bm k}H_{\bm k}\hat{\bm c}^{}_{\bm k},
\end{align}
where
\begin{align}
  \label{eq:Hamiltonian_qsh}
  H_{\bm k}&=(2t-t\cos k_x-t\cos k_y-\mu)s_z\otimes\rho_0\notag\\
  &\quad +a\sin k_x s_z\otimes\rho_x+a\sin k_y s_0\otimes\rho_y,\\
  \hat{\bm c}^{}_{\bm k}&=(\hat{c}^{}_{A,\uparrow{\bm k}},\hat{c}^{}_{B,\uparrow{\bm k}},\hat{c}^{}_{A,\downarrow{\bm k}},\hat{c}^{}_{B,\downarrow{\bm k}})^{\top}.
\end{align}
Here, $s_i$ and $\rho_i$ are the Pauli matrices corresponding to the spin degrees of freedom ($\uparrow, \downarrow$) and the sublattice degrees of freedom ($A, B$), respectively.
This Hamiltonian has only time-reversal symmetry $T$, and we can add perturbations that preserve time-reversal symmetry.
The band structure of the perturbed system is shown in Fig.~\ref{fig:qsh}(a), which reveals the presence of linearly dispersing edge modes.

We compute the optical conductivity by diagonalizing the perturbed Hamiltonian using the expression~\eqref{eq:sigma_band}.
The result is shown in Fig.~\ref{fig:qsh}(b).
These results show that the ${\mathbb Z}_2$ topological insulator does not exhibit the optical conductivity in the low-energy region.
We then compute the spatially-resolved optical conductivity using the recursive Green's function method on the expression~\eqref{eq:sigma_Green}.
The results are presented in Fig.~\ref{fig:qsh}(c).
The results indicate that the spatially-resolved optical conductivity of the ${\mathbb Z}_2$ topological insulator remains constant and does not depend on $\omega$ in the low energy regime.
This behavior is similar to that of the Chern insulator~\cite{James_Majorana}.

To understand these behaviors, we discuss an effective edge theory~\cite{Murakami-Nagaosa_spin_Hall}.
(See Supplemental Material for the detailed calculations.)
The edge modes of a ${\mathbb Z}_2$ topological insulator possess only time-reversal symmetry $T$.
The Hamiltonian allowed under the constraint of time-reversal symmetry is
\begin{align}
  \hat{H}^{\rm edge}&=\sum_{k_x}\hat{\bm f}^{\dagger}_{k_x}H^{\rm edge}_{k_x}\hat{\bm f}^{}_{k_x}
  ,\\
  \label{eq:qsh_edge_Hamiltonian}
  H^{\rm edge}_{k_x}&=k_x
  \begin{pmatrix}
    a_z & a_x-\ii a_y\\
    a_x+\ii a_y & -a_z
  \end{pmatrix}.
\end{align}
Here, $\hat{\bm f}^{}_{k}=(\hat{f}^{}_{\uparrow k}, \hat{f}^{}_{\downarrow k})^{\top}$ is a fermionic operator, and $a_i~(i=x,y,z)$ are real parameters.
Because the effective edge Hamiltonian of the insulator has $\text{U}(1)$ symmetry, we can directly couple the edge Hamiltonian to a $\text{U}(1)$ gauge field unlike the superconductor.
Thus, the current operator is
\begin{align}
  \hat{j}^{\rm edge}&=\sum_{k}\hat{\bm f}^{\dagger}_{k_x+q}j^{\rm edge}_{k_x,q}\hat{\bm f}^{}_{k_x},\\
  \label{eq:qsh_edge_current}
  j_{k,q}^{\rm edge}&=\left(1-\frac{\ii q}{2}\right)
  \begin{pmatrix}
    a_z & a_x-\ii a_y\\
    a_x+\ii a_y & -a_z
  \end{pmatrix}.
\end{align}
We find that the Hamiltonian~\eqref{eq:qsh_edge_Hamiltonian} and the current operator~\eqref{eq:qsh_edge_current} can be diagonalized simultaneously.
This implies that the system permits only intra-band excitations, resulting in ${\rm Re}[\sigma^{\rm edge}(\omega,q=0)]=0$.
Ref.~\cite{Han-James} also points out that the optical conductivity vanishes in the normal helical edge state.
On the other hand, for the momentum-dependent optical conductivity at $T=0$, we have
\begin{align}
    \label{eq:sigma_qsh}
  &\quad{\rm Re}[\sigma^{\rm edge}(\omega,q)]_{T=0}\notag\\
  &=\left(1+\frac{q^2}{4}\right)\left\{\delta\left(q-\frac{\omega}{a}\right)+\delta\left(q+\frac{\omega}{a}\right)\right\},
\end{align}
where $a_0=\sqrt{a_x^2+a_y^2+a_z^2}$.
From Eq.~\eqref{eq:sigma_Fourier}, the spatially-resolved optical conductivity in the low energy regime is
\begin{align}
    \label{eq:sigma_q0_qsh}
  {\rm Re}[\tilde{\sigma}^{\rm edge}(\omega,R_x)]_{T=0}\sim\frac{1}{2\pi^2},
\end{align}
where $\omega\ll a_0$.
As shown in Fig.~\ref{fig:helical}(c), this value is consistent with the numerical simulation.

\subsection{Strong topological superconductor}
\label{sec:helical}
\begin{figure}[!t]
	\centering
	\includegraphics[width=0.95\columnwidth]{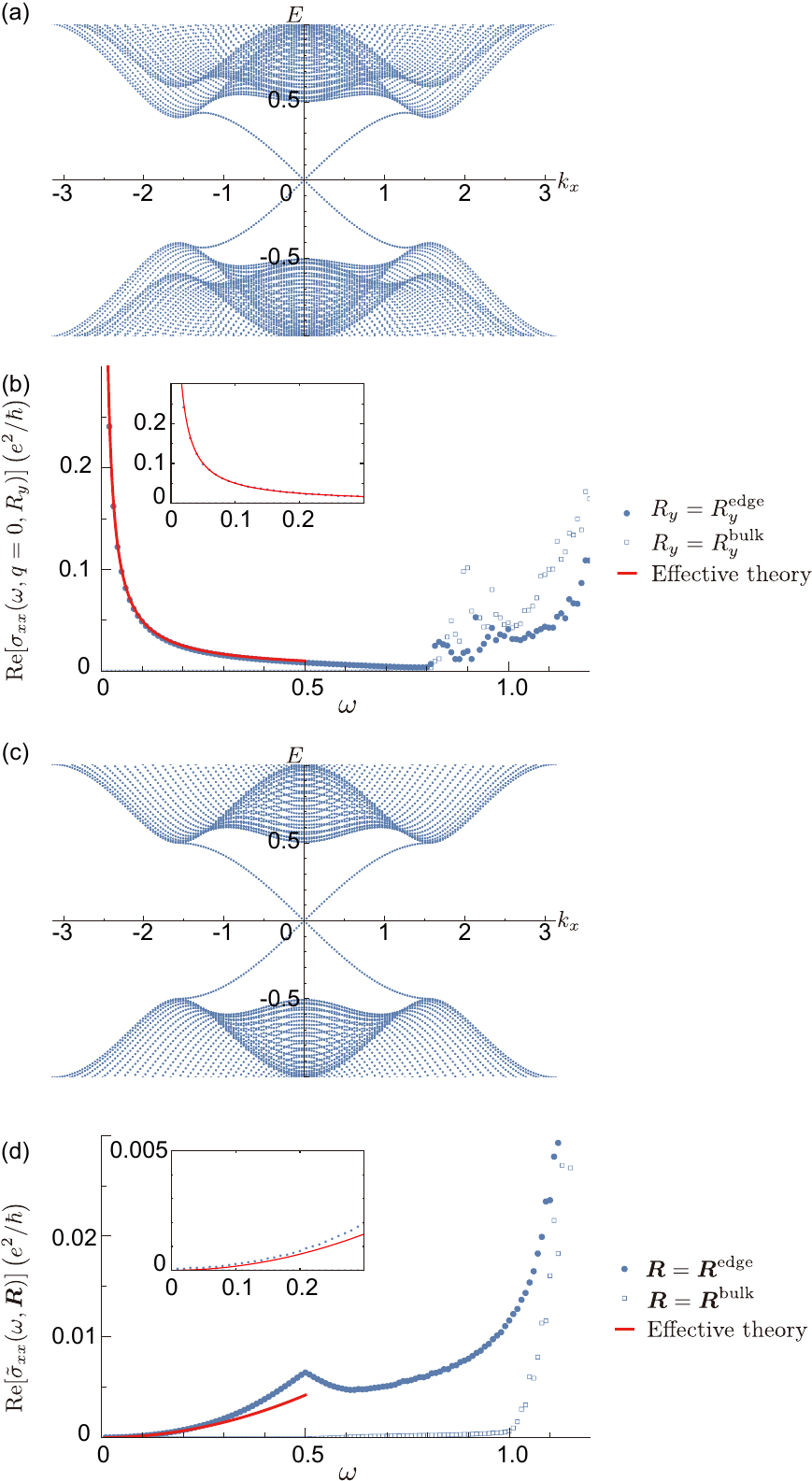}
	\caption{
		(a) The band structure of the strong topological superconductor with $\nu=0.1$. The other parameters used are: $t=\mu=1$ and $\Delta=0.5$. 
        (b) Results of numerical calculation for the optical conductivity based on Eq.~\eqref{eq:sigma_band} at $T=10^{-3}$ and Eq.~\eqref{eq:sigma_q0_helical} at $T=0$. 
        The choice of $R_y$ is the same as in Fig.~\ref{fig:optical_conductivity}.
        (c) The band structure of the strong topological superconductor with $\nu=0$. The other parameters used are: $t=\mu=1$ and $\Delta=0.5$. 
        (d) Results of numerical calculation for the spatially-resolved optical conductivity with $\nu=0$ and $X=1$, based on Eq.~\eqref{eq:sigma_Green} at $T=10^{-3}$ and Eq.~\eqref{eq:sigma_helical} at $T=0$.
        The choice of ${\bm R}$ is the same as in Fig.~\ref{fig:local_optical_conductivity}.
	}
	\label{fig:helical}
\end{figure}

As a generalization of results in Ref.~\cite{James_Majorana} to the time-reversal symmetric case, we consider a two-dimensional strong topological superconductor~\cite{Periodic_table_PRB, Kitaev_bott, Ryu_2010}.
The representative model of this phase is obtained by stacking $p+\ii p$ and $p-\ii p$ superconductors~\cite{Periodic_table_PRB}.
The normal-phase Hamiltonian and superconducting order parameter of this system are given by
\begin{align}
  h_{{\bm k}}&=(\mu-t\cos k_x-t \cos k_y)s_0+\nu \sin k_x s_y,\\
  \Delta_{\bm k}&=\Delta \sin k_x s_z+\ii\Delta\sin k_y s_{0}.
\end{align}
Here, $s_i$ are the Pauli matrices corresponding to the spin degrees of freedom.
The parameter $\nu$ represents the hopping between the $p+\ii p$ and $p-\ii p$ superconductors.
This system has time-reversal symmetry $T$, particle-hole symmetry $C$, and chiral symmetry $S$.
The band structure of this system is shown in Fig.~\ref{fig:helical}(a,c), revealing the presence of linearly dispersing edge modes.

Applying the discussion in Sec.~\ref{sec:q0} to edge modes, we examine whether the lowest energy optical excitation with $q=0$ is allowed or not.
Suppose that the system does not have spatial symmetries other than translations. 
Since $T$ and $C$ transform ${\bm k}$ to $-{\bm k}$, only $S$ is a symmetry of the edge. 
As a result, the EAZ class is AIII, and thus, this system can exhibit the optical conductivity in the low-energy region as shown in Table~\ref{table:EAZ_class}.

We compute the optical conductivity with the expression~\eqref{eq:sigma_band} based on diagonalization, which is shown in Fig.~\ref{fig:helical}(b). 
This numerical result shows that the optical conductivity exists in the low energy regime.
In contrast, the ${\mathbb Z}_2$ topological insulator does not exhibit the optical conductivity, allowing for a clear distinction between it and the strong topological superconductor.

Next, we use the effective edge theory to understand the behavior of the optical conductivity in this system.
(For the detailed calculations, see Supplemental Material.)
The projection onto the effective theory is given by
\begin{align}
  \chi=\frac{1}{\sqrt{2}}
  \begin{pmatrix}
    1 & 0\\
    0 & 1\\
    1 & 0\\
    0 & 1
  \end{pmatrix},
\end{align}
and the effective Hamiltonian and edge current operator are given by
\begin{align}
  H^{\rm edge}_{k_x}&=\Theta(-k_c\leq k_x\leq k_c)
  \begin{pmatrix}
    \Delta k_x & 0\\
    0 & -\Delta k_x
  \end{pmatrix}
  ,\\
  j^{\rm edge}_{k_x,q}&=\Theta(-k_c\leq k_x\leq k_c)\Theta(-k_c\leq k_x+q\leq k_c)\notag\\
  &\qquad \times
  \begin{pmatrix}
    t(k_x+\tfrac{q}{2}) & -\ii \nu(1-\ii \frac{q}{2})\\
    \ii \nu(1-\ii\frac{q}{2}) & t(k_x+\tfrac{q}{2})
  \end{pmatrix}
  .
\end{align}
Here, to avoid the fermion doubling problem, we linearize and introduce a momentum cutoff $k_c$ using
\begin{align}
  \Theta(P)=
  \begin{dcases}
    1\qquad (\text{if } P \text{ is true})\\
    0\qquad (\text{otherwise})
  \end{dcases}.
\end{align}
Using this effective theory, the optical conductivity is given by
\begin{align}
    \label{eq:sigma_q0_helical}
  &\quad {\rm Re}[\sigma^{\rm edge}(\omega,q=0)]_{T=0}\notag\\
  &=
  \begin{dcases}
    \frac{\nu^2}{4\Delta\omega}
    \quad(0\leq\omega\leq 2\Delta k_c)\\
    0\qquad(\omega > 2\Delta k_c)
  \end{dcases}.
\end{align}
At $q=0$, the contribution from intraband excitations becomes zero, so the optical conductivity arises solely from interband excitations.
A more general extension of these results has already been given in Ref.~\cite{Han-James}.
We can confirm in Fig.~\ref{fig:helical}(b) that this result is consistent with the numerical calculation using diagonalization.

On the other hand, when $\nu=0$, the contribution from interband excitations vanishes, leading to ${\rm Re}[\sigma^{\rm edge}(\omega,q=0)]_{T=0}=0$.
This is because, an additional mirror symmetry $P=\ii s_y\otimes\tau_{0}$ exists when $\nu=0$. The products $PC$ and $PT$ become $k$-local operators, with $-(PC)^2=(PT)^2=1$. As a result, the edge modes belong to class CI, prohibiting $q=0$ excitations.
Therefore, in this case, it is important to consider the spatially-resolved optical conductivity.
We compute the spatially-resolved optical conductivity using the recursive Green's function method based on expression~\eqref{eq:sigma_Green}, and the results are shown in Fig.~\ref{fig:helical}(d).
From these results, we observe that the spatially-resolved optical conductivity of the strong topological superconductor shows $\omega$ dependence in the low energy regime.
In contrast, the spatially-resolved optical conductivity of the ${\mathbb Z}_2$ topological insulator is almost constant in the low energy regime, independent of $\omega$, highlighting a significant difference between the two.
Therefore, we can distinguish between the two based on the spatially-resolved optical conductivity even when $\nu=0$.
To analyze this spatially-resolved optical conductivity, we again use the effective edge theory.
Under the condition $\nu=0$, the momentum-dependent optical conductivity is given by
\begin{align}
    \label{eq:sigma_helical}
  &\quad {\rm Re}[\sigma^{\rm edge}(\omega,q)]_{T=0}\notag\\
  &=\frac{t^2\omega^2}{48\Delta^3}\left\{\delta(\omega-\Delta q)+\delta(\omega+\Delta q)\right\}\notag\\
  &\qquad \times
  \begin{dcases}
    1
        \quad(0\leq\omega\leq\Delta k_c)\\
    \left(\frac{2\Delta k_c}{\omega}-1\right)^3
    \quad(\Delta k_c\leq\omega\leq2\Delta k_c)\\
    0\qquad(\omega>2\Delta k_c)
  \end{dcases}.
\end{align}
This result represents a simple sum of contributions from the chiral edge modes of the $p+\ii p$ and $p-\ii p$ superconductors~\cite{James_Majorana}.
We show the comparison between the spatially-resolved optical conductivity obtained from the edge effective theory and the numerical results from the recursive Green's function method in Fig.~\ref{fig:helical}(d).
As $\omega$ increases, there is a discrepancy between the two results, which is due to the deviation of the edge modes from perfect linear dispersion.
However, the two results are consistent with each other.

\subsection{Topological crystalline superconductors in layer group $pmaa$}
\label{sec:pmaa}
\begin{figure}[t]
	\centering
	\includegraphics[width=0.95\columnwidth]{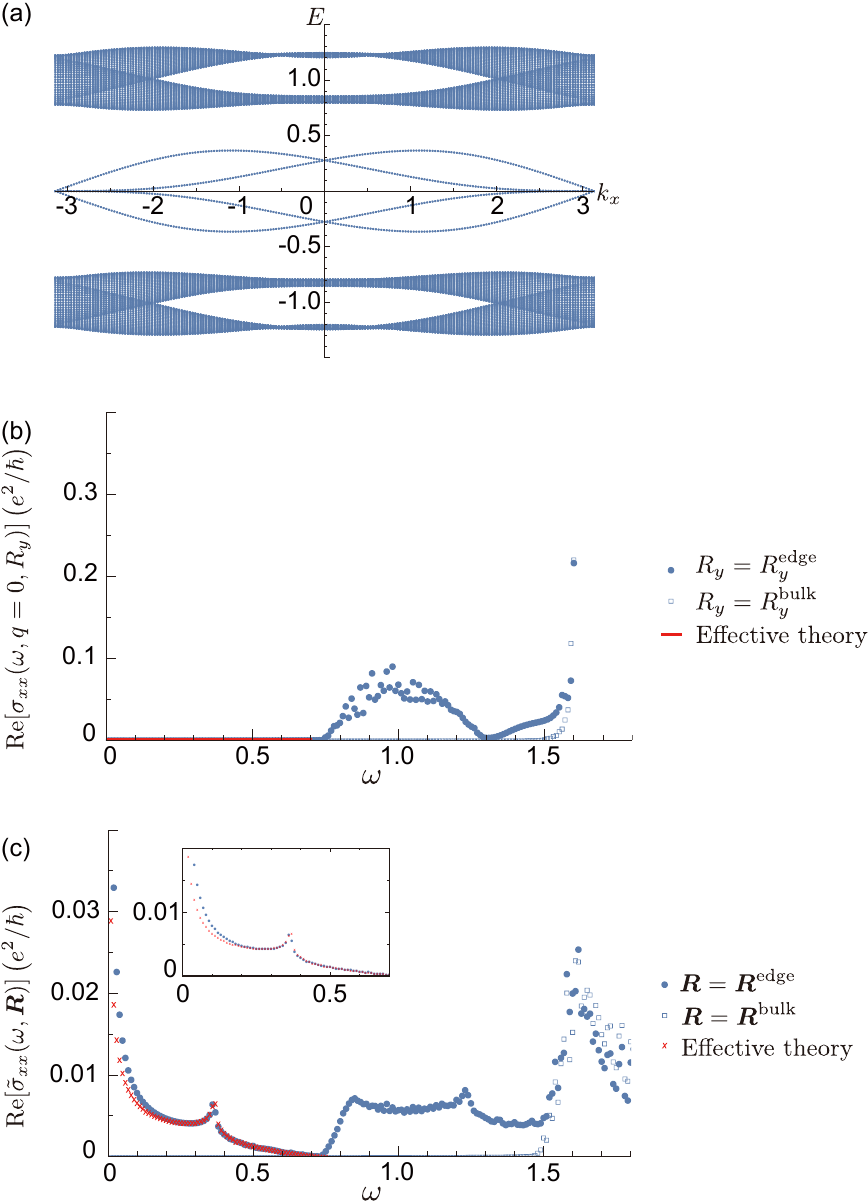}
	\caption{
        (a) The band structure of topological crystalline superconductors belonging to the layer group $pmaa$. The parameters are $t_y=\Delta=1$, $t_x=0.1$, and $t_1=t_2=m_1=m_2=m_3=m_4=0.1$. 
        (b) Results of numerical calculation for the optical conductivity based on Eq.~\eqref{eq:sigma_band} at $T=10^{-3}$. 
        The choice of $R_y$ is the same as in Fig.~\ref{fig:optical_conductivity}.
        (c) Results of numerical calculation for the spatially-resolved optical conductivity with $X=1$, based on Eq.~\eqref{eq:sigma_Green} and the edge effective theory.
        (See Supplemental Material for the expressions.)
        The choice of ${\bm R}$ is the same as in Fig.~\ref{fig:local_optical_conductivity}.
	}
	\label{fig:pmaa}
\end{figure}

We consider topological crystalline superconductors as time-reversal symmetric topological superconductors other than strong topological superconductors.
These systems can be constructed by stacking 1D topological superconductors in a manner that satisfies crystalline symmetries~\cite{defect_network, R-AHSS_Shiozaki, R-AHSS_Song, Shiozaki-Ono2023, Wire_Fang, Ono-Shiozaki-Watanabe2022}.
Here, we focus on a topological crystalline superconductor in layer group $pmaa$.
The normal-phase Hamiltonian and superconducting order parameter of this system are given by
\begin{align}
  h_{{\bm k}}&=-\{t_x(\cos k_x-1)+t_y \cos k_y\}\tau_0\otimes s_0\notag\\
  &\quad +t_1\sin k_x\tau_z\otimes s_x+t_2
  \begin{pmatrix}
    0 & 1+e^{-\ii k_x}\\
    1+e^{\ii k_x} & 0
  \end{pmatrix}
  \otimes s_{0}
  ,\\
  \Delta_{\bm k}&=\ii \Delta \sin k_y \tau_z \otimes s_0+m_1\sin k_x\tau_{z}\otimes s_{z}\notag\\
  &\quad +\ii m_2
  \begin{pmatrix}
    0 & 1+e^{-\ii k_x}\\
    1+e^{\ii k_x} & 0
  \end{pmatrix}
  \otimes s_{y}\notag\\
  &\quad +\ii\{m_3+m_4(\cos k_x-1)\} \tau_{0}\otimes s_{y}.
\end{align}
Here, $s_i$ and $\tau_i$ are the Pauli matrices corresponding to the spin and sublattice degrees of freedom, respectively.
The band structure of this system is shown in Fig.~\ref{fig:pmaa}(a), exhibiting edge modes that are completely decoupled from the bulk bands.

\begin{figure}[t]
	\centering
	\includegraphics[width=0.9\columnwidth]{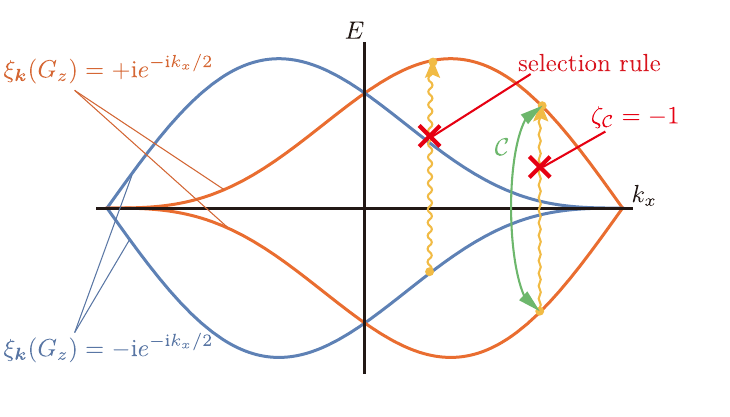}
	\caption{
            The illustration of optical excitation at the edge of a topological crystalline superconductor in the layer group $pmaa$.
            The color of each edge mode corresponds to the glide symmetry eigenvalues.
            Optical excitation between different eigenvalues sectors is suppressed by the selection rule, while optical excitation within the same eigenvalue sector is suppressed by the symmetry $\mathcal{C}$.
            As a result, for topological crystalline superconductors in the layer group $pmaa$, the optical conductivity is absent in the low energy regime.
	}
	\label{fig:pmaa_edge}
\end{figure}
First, we consider the presence or absence of the optical conductivity in the low energy regime from the perspective of symmetry.
The crystalline symmetry group $G$ of this system is generated by glide symmetry $G_z$ (half translation along $x$-direction followed by reflection about $z$-direction), mirror symmetry $M_x$ along $x$-direction, inversion $I$, and the translation along $y$-direction.
Correspondingly, the symmetry group whose elements remain on the edge is generated by $G_z$ and $M_x$. 
In particular, $G_z$ keeps generic momentum $k_x$ invariant, whose eigenvalues are given by $\xi_{k_x}(G_z)=\pm \ii e^{-\ii k_x/2}$.
In addition to this, the combinations of mirror $M_x$, time-reversal symmetry $T$, and particle-hole symmetry $C$ also give us symmetries ${\mathcal T}=M_xT, {\mathcal C}=M_xC$, and $S=TC$ that do not change generic momentum $k_x$.
As discussed in Sec.~\ref{sec:q0}, suppose that we have an eigenstate ${\bm \psi}_{n{k_x}}$ of $H^{\text{edge}}_{k_x}$ satisfying $U_{k_x}^{\text{edge}}(G_z){\bm \psi}_{n{k_x}}=\xi_{k_x}(G_z){\bm \psi}_{n{k_x}}$.
Then, we ask whether $U_{k_x}^{\text{edge}}({\mathcal T}){\bm \psi}_{n{k_x}}^*, U_{k_x}^{\text{edge}}({\mathcal C}){\bm \psi}_{n{k_x}}^*$, and $U_{k_x}^{\text{edge}}(S){\bm \psi}_{n{k_x}}$ have the same eigenvalue of $G_z$ or not. 
Following the discussions in Appendix~\ref{app:EAZ_class}, we see that they have the same eigenvalue $\xi_{k_x}(G_z)$.
Furthermore, since $\zeta_{\mathcal C}=-\zeta_{\mathcal T}=-1$ in Eqs.~\eqref{eq:T2} and~\eqref{eq:C2}, the EAZ class of each eigensector is class CI. 
From Table~\ref{table:EAZ_class}, it follows that there is no optical response in the low-energy region [see Fig.~\ref{fig:pmaa_edge} for an illustration].
This conclusion is also confirmed by the numerical calculation of the optical conductivity with the expression~\eqref{eq:sigma_band} using diagonalization, as shown in Fig.~\ref{fig:pmaa}(b).

Therefore, in this topological crystalline superconductor, we focus on the spatially-resolved optical conductivity.
We show the numerical calculation results of the spatially-resolved optical conductivity using the recursive Green’s function method based on Eq.~\eqref{eq:sigma_Green} in Fig.\ref{fig:pmaa}(c).
These results show that the spatially-resolved optical conductivity of the topological crystalline superconductor belonging to the layer group $pmaa$ has a strong $\omega$ dependence in the low energy regime.
When compared to the ${\mathbb Z}_2$ topological insulator, both have ${\rm Re}[\sigma_{xx}(\omega,q=0)]=0$ in the low energy regime.
However, the strong $\omega$ dependence of the spatially-resolved optical conductivity is a significant difference.
Moreover, this $\omega$ dependence of the spatially-resolved optical conductivity is also different from that of the strong topological superconductor.
Thus, based on the behavior of the spatially-resolved optical conductivity, it is possible to distinguish between these systems.

Finally, to understand the low energy behavior of the spatially-resolved optical conductivity in the topological crystalline superconductor belonging to the layer group $pmaa$, we discuss the effective edge theory.
When $t_i$ and $m_i$ are sufficiently small, the projection onto the edge modes of this system is given by
\begin{align}
  \chi=\frac{1}{\sqrt{2}}
  \begin{pmatrix}
    1 & 0 & 0 & 0 \\
    0 & 1 & 0 & 0 \\
    0 & 0 & -\ii & 0 \\
    0 & 0 & 0 & -\ii \\
    1 & 0 & 0 & 0 \\
    0 & 1 & 0 & 0 \\
    0 & 0 & \ii & 0 \\
    0 & 0 & 0 & \ii
  \end{pmatrix}.
\end{align}
Then the edge Hamiltonian is given by
\begin{align}
  H^{\rm edge}_{k_x}&=
  \sin k_x
  \begin{pmatrix}
      1 & 0\\
      0 & 0
  \end{pmatrix}
  \otimes
  \begin{pmatrix}
      m_1 & t_1\\
      t_1 & -m_1
  \end{pmatrix}
  \notag\\
  &\quad +\sin k_x
  \begin{pmatrix}
      0 & 0\\
      0 & 1
  \end{pmatrix}
  \otimes
  \begin{pmatrix}
      m_1 & -t_1\\
      -t_1 & -m_1
  \end{pmatrix}\notag\\
  &\quad+(1+e^{-\ii k_x})
  \begin{pmatrix}
      0 & -\ii\\
      0 & 0
  \end{pmatrix}
  \otimes
  \begin{pmatrix}
      t_2 & -m_2\\
      m_2 & t_2
  \end{pmatrix}\notag\\
  &\quad+(1+e^{\ii k_x})
  \begin{pmatrix}
      0 & 0\\
      \ii & 0
  \end{pmatrix}
  \otimes
  \begin{pmatrix}
      t_2 & m_2\\
      -m_2 & t_2
  \end{pmatrix}
\end{align}
and the edge current operator is given by
\begin{align}
  j^{\rm edge}_{k_x,q}&=-t_x e^{-\ii \frac{q}{2}}\sin \left(k_x+\frac{q}{2}\right)
  \begin{pmatrix}
    1 & & & \\
    & 1 & & \\
    & & 1 & \\
    & & & 1
  \end{pmatrix}.
\end{align}

The optical conductivity calculation $\sigma^{\rm edge}(\omega,q)$ using this effective theory is generally complicated.
(See Supplemental Material for the detailed calculations and expressions.)
We show the comparison between the numerical results obtained using the recursive Green's function and those calculated using the effective theory in Fig.~\ref{fig:pmaa}(d).
In the low energy regime, there are slight errors due to the influence of the convergence factor $\eta$, but the two results are consistent.

\subsection{Topological crystalline superconductors in layer group $p11a$}
\label{sec:p11a}
\begin{figure}[t]
	\centering
	\includegraphics[width=0.95\columnwidth]{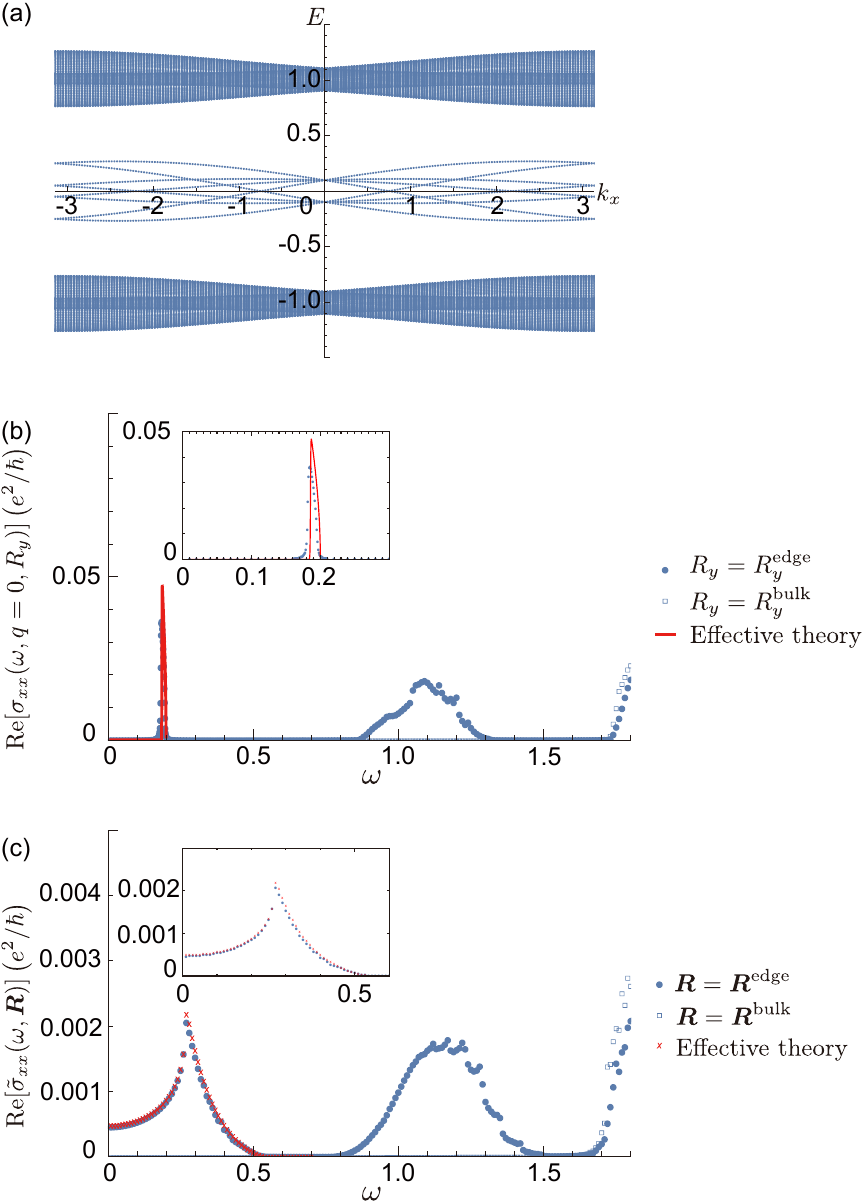}
	\caption{
	(a) The band structure of topological crystalline superconductors belonging to the layer group $p11a$. The parameters are $t=\Delta=1$, $t_1 = 0.05$, $t_2 = -0.05$, $t_3 = 0$, $t_4 = 0$, $m_1 = 0$, $m_2 = -0.05$, $m_3 = -0.075$, and $m_4 = 0.075$. 
        (b) Results of numerical calculation for the optical conductivity based on Eq.~\eqref{eq:sigma_band} and Eq.~\eqref{eq:sigma_q0_p11a} at $T=10^{-3}$. 
        The choice of $R_y$ is the same as in Fig.~\ref{fig:optical_conductivity}.
        (c) Results of numerical calculation for the spatially-resolved optical conductivity with $X=1$, based on Eq.~\eqref{eq:sigma_Green} and the edge effective theory at $T=10^{-3}$.
        (See Supplemental Material for the expressions.)
        The choice of ${\bm R}$ is the same as in Fig.~\ref{fig:local_optical_conductivity}.
	}
	\label{fig:p11a}
\end{figure}
As another example of topological crystalline superconductors, we consider a tight-binding model with layer group $p11a$~\cite{Shiozaki-Sato_crystalline_insulator_SC,Shiozaki-Sato-Gomi2016,Shiozaki-Sato-Gomi_mobius_twist}, which is a subgroup of layer group $pmaa$.
The normal part and pairing function of the Hamiltonian for this system are given by
\begin{align}
    h_{\bm k}&=-t\cos k_y \rho_{0}\otimes s_{0}
    +t_1
    \begin{pmatrix}
        0 & 1+e^{-\ii k_x}\\
        1+e^{\ii k_x} & 0
    \end{pmatrix}
    \otimes s_0
		\notag\\
    &\quad +\ii t_2
    \begin{pmatrix}
        0 & 1-e^{-\ii k_x}\\
        -1+e^{\ii k_x} & 0
    \end{pmatrix}
    \otimes s_z
    \notag\\
    &\quad
    +\ii
    \begin{pmatrix}
        0 & 1+e^{-\ii k_x}\\
        -1-e^{\ii k_x} & 0
    \end{pmatrix}
    \otimes \left(t_3s_x+t_4s_y\right),
		\\
    \Delta_{\bm k}&=\Delta\sin k_y \rho_0\otimes s_z
    +\ii m_1
    \begin{pmatrix}
        0 & 1-e^{-\ii k_x}\\
        1-e^{\ii k_x} & 0
    \end{pmatrix}
    \otimes s_y
		\notag\\
    &\quad +\ii m_2
    \begin{pmatrix}
        0 & 1+e^{-\ii k_x}\\
        -1-e^{\ii k_x} & 0
    \end{pmatrix}
    \otimes s_x
    \notag\\
    &\quad
    +\ii
    \begin{pmatrix}
        0 & 1-e^{-\ii k_x}\\
        -1+e^{\ii k_x} & 0
    \end{pmatrix}
    \otimes \left(m_3s_0+m_4s_z\right).
\end{align}
Here, $\rho_i$ and $s_i$ are the Pauli matrices corresponding to sublattice and spin degrees of freedom, respectively.

\begin{figure}[t]
	\centering
	\includegraphics[width=0.9\columnwidth]{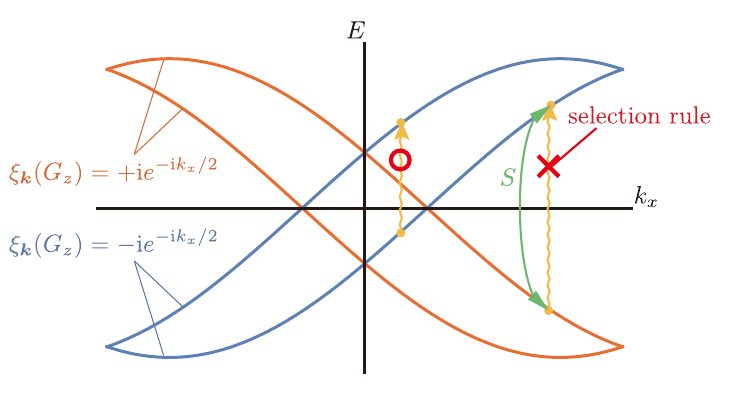}
	\caption{
            The illustration of optical excitation at the edge of a topological crystalline superconductor in the layer group $p11a$.
            The color of each edge mode corresponds to the glide symmetry eigenvalues.
            Optical excitation between different eigenvalue sectors is suppressed by the selection rule, while optical excitation within the same eigenvalue sector is allowed.
            Therefore, for topological crystalline superconductors in the layer group $p11a$, the optical excitation is allowed in the low energy regime.
	}
	\label{fig:p11a_edge}
\end{figure}
Again, we determine whether optical excitation in the low energy regime is allowed or not based on symmetries.
The crystalline symmetry group $G$ is generated by the glide symmetry $G_z$ and translation symmetry along $y$-direction.
On the edge, the crystalline symmetry group $G^{\text{edge}}$ is generated by $G_z$. 
As in the case of the layer group $pmaa$, $G_z$ does not change generic momentum ${\bm k}$, and its eigenvalues are $\xi_{\bm k}(G_z)=\pm\ii e^{-\ii k_x/2}$.
On the other hand, unlike the layer group $pmaa$, there is no crystalline symmetry that transforms $k_x$ to $-k_x$, which implies that neither ${\mathcal T}$ nor ${\mathcal C}$ exists. 
Since $S = CT$ is always an onsite symmetry, we consider the eigenvalue of $G_z$ for $U^{\text{edge}}(S){\bm \psi}_{n{k_x}}$.
From the discussions in Appendix~\ref{app:EAZ_class}, when an eigenstate ${\bm \psi}_{n{k_x}}$ with the highest negative energy has eigenvalue $\xi_{\bm k}(G_z)$, we find that $U^{\text{edge}}(S){\bm \psi}_{n{k_x}}$ has eigenvalue $-\xi_{\bm k}(G_z)$.
This indicates that the lowest-energy optical excitation is still prohibited, since any optical excitation cannot happen between eigenstates with different eigenvalues. 
However, due to the glide symmetry and nontrivial topology, the number of edge modes is four on the edge. 
Thanks to the presence of multibands, optical excitation between ${\bm \psi}_{n{k_x}}$ and an eigenstate other than $U^{\text{edge}}(S){\bm \psi}_{n{k_x}}$ is allowed in low energy regime [see Fig.~\ref{fig:p11a_edge} for an illustration].
We show the results of the numerical calculation of the optical conductivity using expression~\eqref{eq:sigma_band} based on diagonalization in Fig.~\ref{fig:p11a}(b).
This numerical result is consistent with the conclusion derived from the above symmetry argument.

To understand the behavior of this optical conductivity, we use an effective edge theory.
See Supplemental Material for the detailed calculations and expressions.
When $t_i$ and $m_i$ are sufficiently small, the projection onto the edge modes of this system is given by
\begin{align}
    \chi&=\frac{1}{\sqrt{2}}
    \begin{pmatrix}
    e^{-\ii\frac{\pi}{4}} & 0 & 0 & 0\\
    0 & e^{\ii\frac{\pi}{4}} & 0 & 0\\
    0 & 0 & e^{-\ii\frac{\pi}{4}} & 0\\
    0 & 0 & 0 & e^{\ii\frac{\pi}{4}}\\
    e^{\ii\frac{\pi}{4}} & 0 & 0 & 0\\
    0 & e^{-\ii\frac{\pi}{4}} & 0 & 0\\
    0 & 0 & e^{\ii\frac{\pi}{4}} & 0\\
    0 & 0 & 0 & e^{-\ii\frac{\pi}{4}}\\
  \end{pmatrix}
  .
\end{align}
Then, the edge Hamiltonian is given by
\begin{align}
  H^{\rm edge}_{k_x}&=
  \ii(m_3+t_2)
  \begin{pmatrix}
      0 & 1-e^{-\ii k_x}\\
      1-e^{\ii k_x} & 0
  \end{pmatrix}
  \otimes
  \begin{pmatrix}
      1 & 0\\
      0 & -1
  \end{pmatrix}
  \notag\\
  &\quad +\ii(m_2-t_4)
  \begin{pmatrix}
      0 & -(1+e^{-\ii k_x})\\
      1+e^{\ii k_x} & 0
  \end{pmatrix}
  \otimes
  \begin{pmatrix}
      0 & 1\\
      1 & 0
  \end{pmatrix}
\end{align}
and the edge current operator is given by
\begin{align}
  j^{\rm edge}_{k_x,q}&=\frac{\ii}{2}
  \begin{pmatrix}
      0 & 1-e^{-\ii(k_x+q)}\\
      0 & 0
  \end{pmatrix}
  \otimes
  \begin{pmatrix}
      t_1 & -t_3\\
      t_3 & t_1
  \end{pmatrix}
  \notag\\
  &\quad +\frac{\ii}{2}
  \begin{pmatrix}
      0 & 0\\
      1-e^{\ii k_x} & 0
  \end{pmatrix}
  \otimes
  \begin{pmatrix}
      -t_1 & -t_3\\
      t_3 & -t_1
  \end{pmatrix}
  .
\end{align}

Using this effective edge theory, the optical conductivity is given by
\begin{widetext}
    \begin{align}
    \label{eq:sigma_q0_p11a}
      {\rm Re}[\sigma^{\rm edge}(\omega,q=0)]&=\frac{t_1^2+t_3^2}{2\omega m_0}\frac{1}{\left|\sin\alpha\right|}\sqrt{1-\left(\frac{\omega}{2m_0 \sin\alpha}\right)^2}\notag\\
      &\quad \times\left\{f\left(m_0 \cos\alpha\sqrt{1-\left(\frac{\omega}{2m_0\sin\alpha}\right)^2}-\frac{\omega}{2}\right)-f\left(m_0 \cos\alpha\sqrt{1-\left(\frac{\omega}{2m_0 \sin\alpha}\right)^2}+\frac{\omega}{2}\right)\right.\notag\\
      &\qquad \quad \left. +f\left(-m_0 \cos\alpha\sqrt{1-\left(\frac{\omega}{2m_0\sin\alpha}\right)^2}-\frac{\omega}{2}\right)-f\left(-m_0 \cos\alpha\sqrt{1-\left(\frac{\omega}{2m_0 \sin\alpha}\right)^2}+\frac{\omega}{2}\right)\right\},
    \end{align}
\end{widetext}
where $m_0=2\sqrt{(m_2-t_4)^2+(m_3+t_2)^2}$ and $\alpha=\arctan\left(\frac{m_2-t_4}{m_3+t_2}\right)$.
This result is consistent with the numerical calculation shown in Fig.~\ref{fig:p11a}(b).
Furthermore, it can be seen from Eq.~\eqref{eq:sigma_q0_p11a} that optical conductivity exists in the energy region $\omega \in [m_0 \sin 2\alpha, 2m_0 \sin \alpha]$ at $T=0$.

On the other hand, the spatially-resolved optical conductivity calculated using the recursive Green's function method based on expression~\eqref{eq:sigma_Green} is shown in Fig.~\ref{fig:p11a}(c).
The results indicate that spatially-resolved optical conductivity exists even in the range $\omega\in[0,m_0\sin 2\alpha]$, where optical conductivity is absent.
Moreover, the behavior of this spatially-resolved optical conductivity can be understood using the effective theory as shown in Fig.~\ref{fig:p11a}(c).

\section{Concusion}
\label{sec:conclusion}
In this work, we analyzed the optical response of topological superconductors and compared the optical conductivity and spatially-resolved optical conductivity in topological superconductors with those in a two-dimensional ${\mathbb Z}_2$ topological insulator.
We performed numerical calculations based on the bulk Hamiltonians and analytical calculations using the effective edge theory. 
Here, we summarize our findings. 

In a two-dimensional ${\mathbb Z}_2$ topological insulator, optical conductivity, $\sigma_{xx}(\omega, q=0)$, is absent in the low-energy regime.
Therefore, nonzero optical conductivity in the low energy regime implies topological superconductivity if the bulk is gapped.
For superconductors, whether lowest energy optical excitation is allowed or not can be determined by symmetries.
By combining symmetry analysis, numerical simulations, and analytical calculations, we find that optical conductivity in a low energy regime is present for strong topological superconductors without mirror symmetry and for topological crystalline superconductors with layer group $p11a$. 

When optical conductivity in a low energy regime must be zero due to symmetries, we discuss spatially-resolved optical conductivity, $\sigma_{xx}(\omega, {\bm R})$.
For the two-dimensional ${\mathbb Z}_2$ topological insulator, the spatially-resolved optical conductivity is constant in the low energy regime.
On the other hand, for strong topological superconductors and topological crystalline superconductors, the spatially-resolved optical conductivity generally exhibits $\omega$-dependent behaviors.
These results suggest that even under time-reversal symmetry and crystalline symmetry, edge modes in normal-conducting and superconducting phases exhibit distinct optical responses.

As a more realistic analysis, the momentum-dependent optical conductivity $\sigma(\omega, q)$ derived from the edge effective theory could be convolved with the actual optical system to match experimental setups.

\begin{acknowledgements}
    H.K.~thanks Yohei Fuji, Hosho Katsura, Sota Kitamura, Koki Okajima, Yugo Onishi, Hisanori Oshima, and Sena Watanabe for helpful discussions.
    H.K.~and S.O.~also thank Haruki Watanabe for helpful discussions and for encouraging them to complete this work.
    H.K.~is supported by Forefront Physics and Mathematics Program to Drive Transformation (FoPM), a World-leading Innovative Graduate Study (WINGS) Program, the University of Tokyo.
    J.J.H.~is supported by National Natural Science Foundation of China (Grant No.~12204451).
    S.O.~was supported by KAKENHI Grant No.~JP20J21692 from the Japan Society for the Promotion of Science (JSPS) and RIKEN Special Doctoral Research Program.
\end{acknowledgements}

\clearpage
{
\appendix
\section{Relation between spatially-resolved and momentum dependent optical conductivity}
\label{app:local_momentum_opt_cond}
We derive the relationship between the spatially-resolved optical conductivity $\tilde{\sigma}_{xx}(\omega,{\bm R})$ and the momentum dependent optical conductivity $\sigma_{xx}(\omega,q,R_y)$.
Given the translational symmetry of the system in the $x$ direction, $\tilde{\sigma}_{xx}(\omega,{\bm R})$ can be expressed as
\begin{widetext}
    \begin{align}
    	\tilde{\sigma}_{xx}(\omega,{\bm R})
    	 &=\frac{1}{L_x}\sum_{X_0}\tilde{\sigma}_{xx}(\omega,{\bm R}+X_0{\bm a}_x)\notag\\
    	 &=\frac{1}{2\pi\omega L_x}\sum_{X_0}\int_{0}^{\infty}dt\ e^{\ii (\omega+\ii\eta)t}\langle[e^{\ii\hat{H}t}\hat{J}^{x}({\bm R}+X_0{\bm a}_x)e^{-\ii\hat{H}t},\hat{J}^{x}({\bm R}+X_0{\bm a}_x)]\rangle\notag\\
    	 &=\frac{1}{2\pi\omega L_x X^2}\sum_{X_0}\sum_{X',X''=0}^{X-1}\sum_{Y',Y''=0}^{Y-1}\int_{0}^{\infty}dt\ e^{\ii (\omega+\ii \eta)t}\langle[e^{\ii\hat{H}t}\hat{j}^{x}_{{\bm R}+(X_0+X'){\bm a}_x+Y'{\bm a}_y}e^{-\ii\hat{H}t},\hat{j}^{x}_{{\bm R}+(X_0+X''){\bm a}_x+Y''{\bm a}_y}]\rangle\notag\\
    	 &=\frac{1}{2\pi\omega L_x^3X^2}\sum_{X_0}\sum_{X',X''=0}^{X-1}\sum_{q,q^{\prime}}e^{\ii (q-q^{\prime})X_0}e^{\ii qX'}e^{-\ii q'X''}\int_{0}^{\infty}dt\ e^{\ii (\omega+\ii\eta)t}\langle[e^{\ii\hat{H}t}(\hat{j}^{x}_{q,R_y})^{\dagger}e^{-\ii\hat{H}t},\hat{j}^{x}_{q^{\prime},R_y}]\rangle\notag\\
    	 &=\frac{1}{2\pi\omega L_x^2X^2}\sum_{X',X''=0}^{X-1}\sum_{q}e^{\ii q(X'-X'')}\int_{0}^{\infty}dt\ e^{\ii (\omega+\ii \eta)t}\langle[e^{\ii\hat{H}t}(\hat{j}^{x}_{q,R_y})^{\dagger}e^{-\ii\hat{H}t},\hat{j}_{q,R_y}^x(0)]\rangle\notag\\
    	 &=\frac{1}{2\pi L_xX^2}\sum_{X',X''=0}^{X-1}\sum_{q}e^{\ii q(X'-X'')}\sigma_{xx}(\omega,q,R_y).
    \end{align}
\end{widetext}
This result can be understood as the Fourier transform of $\tilde{\sigma}_{xx}(\omega,{\bm R})$ in the $x$ direction.
Finally, taking the continuous limit
\begin{align}
  \frac{1}{L_x}\sum_{q}\rightarrow \int_{-\pi}^{\pi}\frac{dq}{2\pi},
\end{align}
we obtain the relationship between $\tilde{\sigma}_{xx}(\omega,{\bm R})$ and $\sigma_{xx}(\omega,q,R_y)$:
\begin{align}
  &\quad \tilde{\sigma}_{xx}(\omega,{\bm R})\notag\\
  &=\frac{1}{2\pi X^2}\sum_{X',X''=0}^{X-1}\int_{-\pi}^{\pi} \frac{dq}{2\pi} e^{\ii q\cdot(X'-X'')}\sigma_{xx}(\omega,q,R_y).
\end{align}

\section{symmetry and EAZ class}
\label{app:EAZ_class}
We discuss eigenvalue sectors and EAZ classes based on symmetry.
First, considering the projective representation of the symmetry group $G$ of the system, we have
\begin{align}
  U_{h{\bm k}}(g)U_{\bm k}^{\phi_g}(h)=z_{g,h}U(gh)\quad(g,h\in G).
\end{align}
Here, $\phi_g = +1~(-1)$ corresponds to unitary (anti-unitary) operations, respectively. We introduce the notation $C^{\phi_g=+1} = C$ and $C^{\phi_g=-1} = C^*$ for any classical number, vector, or matrix $C$.
Considering a crystalline symmetry $g$ in the symmetry $G$, the eigenstate $\psi_{n{\bm k}}$ of the Hamiltonian is also an eigenstate of $U(g)$, satisfying
\begin{align}
  U_{\bm k}(g){\bm \psi}_{n{\bm k}} = \xi_{{\bm k}}(g){\bm \psi}_{n{\bm k}},
\end{align}
and the states of the system are divided into eigenvalue sectors based on this eigenvalue $\xi_{{\bm k}}(g)$.
By considering the transformations of the symmetry operators ${\mathcal C}$, ${\mathcal T}$, and $S$ within these eigenvalue sectors, we can determine the EAZ class.
The eigenstates ${\bm \psi}_{n{\bm k}}$ within each eigenvalue sector are transformed into $U(a){\bm \psi}_{n{\bm k}}^{\phi_a}$ by the symmetries $a = {\mathcal C}, {\mathcal T}, S$.
Using $h$ such that $ga = ah$, we have
\begin{align}
  U_{\bm k}(g)[U(a){\bm \psi}_{n{\bm k}}^{\phi_a}]&=z_{g,a}U_{\bm k}(ga){\bm \psi}_{n{\bm k}}^{\phi_a}\notag\\
  &=z_{g,a}U_{\bm k}(ah){\bm \psi}_{n{\bm k}}^{\phi_a}\notag\\
  &=\frac{z_{g,a}}{z_{a,h}}U_{\bm k}(a)[U_{\bm k}(h){\bm \psi}_{n{\bm k}}]^{\phi_a}\notag\\
  &=\frac{z_{g,a}}{z_{a,h}}[\xi_{\bm k}(h)]^{\phi_a}[U_{\bm k}(a){\bm \psi}_{n{\bm k}}^{\phi_a}].
\end{align}
Thus, the eigenvalue $\xi'_{\bm k}(g)$ of $U(a){\bm \psi}_{n{\bm k}}^{\phi_a}$ with respect to $U_{\bm k}(g)$ becomes
\begin{align}
    \label{eq:projective_factor_transform}
  \xi'_{\bm k}(g)=\frac{z_{g,a}}{z_{a,h}}[\xi_{\bm k}(h)]^{\phi_a}.
\end{align}
If this transformed eigenvalue $\xi'_{\bm k}(g)$ differs from the initial eigenvalue $\xi_{\bm k}(g)$, the eigenvalue sector is not closed under the symmetry $a$.
On the other hand, if $\xi'_{\bm k}(g) = \xi_{\bm k}(g)$, the eigenvalue sector is closed under the symmetry $a$, and the EAZ class can be determined by considering $\zeta_a = z_{a,a}$.

\section{Edge symmetry}
\label{app:edge_symmetry}
We consider projecting the representation $U_{k_x}(g)$ of a crystalline symmetry $g~\in G^{\rm edge}$.
In the bulk, the representation $U_{k_x}(g)$ of the symmetry $g$ and the Hamiltonian $H_{\bm k}$ satisfy the relation
\begin{align}
  U_{k_x}(g)H_{\bm k}=H_{g{\bm k}}U_{k_x}(g).
\end{align}
When the Hamiltonian is separable as $H_{\bm k}=h^x(k_x)+h^y(k_y)$, each component transforms according to
\begin{align}
  U_{k_x}(g)h^x(k_x)&=h^x(gk_x)U_{k_x}(g),\\
  \label{eq:trsf_y_comp}
  U_{k_x}(g)h^y(k_y)&=h^y(k_y)U_{k_x}(g).
\end{align}
To project the representation $U_{k_x}(g)$ of the symmetry $g$ onto the edge, we first find the basis of the edge modes localized at the edge of interest, $\phi_i(y) \propto \varphi_i(y){\bm \chi}_i~(i=1,\cdots,l')$.
As explained in Sec.~\ref{sec:edge}, we can find this basis by solving the equation for the zero mode in the $y$ direction:
\begin{align}
  \label{eq:y_zero_mode_app}
  h^{y}(-i\partial_y)\varphi_i(y){\bm \chi}_i={\bm 0}.
\end{align}
With the matrix $\chi=({\bm \chi}_1,\cdots,{\bm \chi}_{l'})$ constructed from this basis, we project the Hamiltonian and representation onto the edge as
\begin{align}
  \label{eq:Hedge}
  H^{\rm edge}_{k_x}&=\chi^{\dagger}h^x(k_x)\chi,\\
  \label{eq:Uedge}
  U^{\rm edge}_{k_x}(g)&=\chi^{\dagger}U_{k_x}(g)\chi.
\end{align}
By applying Eq.~\eqref{eq:trsf_y_comp} to Eq.~\eqref{eq:y_zero_mode_app}, we obtain
\begin{align}
  h^{y}(-i\partial_y)f(y)[U^{}_{k_x}(g){\bm \chi}_i]={\bm 0}.
\end{align}
Since $g$ preserves the edge, we can express $U^{}_{k_x}(g){\bm \chi}_i$ as a linear combination of the edge modes ${\bm \chi}_i~(i=1,\cdots,l')$ at the edge of interest.
Thus, we use a unitary matrix $W_{k_x}(g)$ of size $l' \times l'$ to write
\begin{align}
  \label{eq:sewing}
  U_{k_x}(g)\chi=\chi W_{k_x}(g).
\end{align}
Taking the Hermitian conjugate of both sides gives
\begin{align}
  \label{eq:sewing_hermitian}
  \chi^{\dagger}U_{k_x}^{\dagger}(g)=W_{k_x}^{\dagger}(g)\chi^{\dagger}.
\end{align}
Using these relations along with Eq.~\eqref{eq:Hedge} and Eq.~\eqref{eq:Uedge}, we find
\begin{align}
  U^{\rm edge}_{k_x}(g)H^{\rm edge}_{k_x}&=\chi^{\dagger}U_{k_x}(g)\chi\chi^{\dagger}h^x(k_x)\chi\notag\\
  &=\chi^{\dagger}\chi W_{k_x}(g)\chi^{\dagger}h^x(k_x)\chi\notag\\
  &=\chi^{\dagger}U_{k_x}(g)h^x(k_x)\chi\notag\\
  &=\chi^{\dagger}h^x(gk_x)U_{k_x}(g)\chi\notag\\
  &=\chi^{\dagger}h^x(gk_x)\chi W_{k_x}(g)\notag\\
  &=\chi^{\dagger}h^x(gk_x)\chi \chi^{\dagger} U_{k_x}(g)\chi\notag\\
  &=H^{\rm edge}_{gk_x}U^{\rm edge}_{k_x}(g).
\end{align}
Therefore, we conclude that
\begin{align}
  U^{\rm edge}_{k_x}(g)H^{\rm edge}_{k_x}=H^{\rm edge}_{gk_x}U^{\rm edge}_{k_x}(g),
\end{align}
which confirms that $U^{\rm edge}_{k_x}(g)$ is indeed a representation of the symmetry of $H^{\rm edge}_{k_x}$.

\section{Symmetry and optical excitation}
\label{app:symmetry_excitation}
We review the relationship between the symmetry ${\mathcal C}$ and optical excitations~\cite{Ahn-Nagaosa_NC}, and extend the discussion to edge modes.
Under the ${\mathcal C}$ symmetry, the current operator transforms as
\begin{align}
&\quad U_{\bm k}({\mathcal C})(j^i_{\bm k})^*U_{\bm k}^{\dagger}({\mathcal C}) = j^i_{\bm k}.
\end{align}
(For the derivation, see Supplemental Material)
We then consider the matrix element of the current operator between the states ${\bm \psi}_{n{\bm k}}$ and $U_{\bm k}({\mathcal C}){\bm \psi}_{n{\bm k}}$:
\begin{align}
  &\quad {\bm \psi}_{n{\bm k}}^{\dagger}j^{i}_{\bm k}(U_{{\bm k}}({\mathcal C}){\bm \psi}_{n{\bm k}}^*)\notag\\
  &=(j^{i\dagger}_{\bm k}{\bm \psi}_{n{\bm k}})^{\dagger}(U_{{\bm k}}({\mathcal C}){\bm \psi}_{n{\bm k}}^*)\notag\\
  &=\{(U_{{\bm k}}({\mathcal C}){\bm \psi}_{n{\bm k}}^*)^{\dagger}(j^{i}_{\bm k}{\bm \psi}_{n{\bm k}}^{})\}^*\notag\\
  &=(U_{{\bm k}}^*({\mathcal C}){\bm \psi}_{n{\bm k}})^{\dagger}j^{i*}_{\bm k}{\bm \psi}_{n{\bm k}}^{*}\notag\\
  &=(U_{{\bm k}}^*({\mathcal C}){\bm \psi}_{n{\bm k}})^{\dagger}U_{{\bm k}}^{\dagger}({\mathcal C})U_{{\bm k}}({\mathcal C})j^{i*}_{\bm k}U_{{\bm k}}^{\dagger}({\mathcal C})U_{{\bm k}}({\mathcal C}){\bm \psi}_{n{\bm k}}^{*}\notag\\
  &=(U_{{\bm k}}^{}({\mathcal C})U_{{\bm k}}^*({\mathcal C}){\bm \psi}_{n{\bm k}})^{\dagger}\{U_{{\bm k}}({\mathcal C})j^{i*}_{\bm k}U_{{\bm k}}^{\dagger}({\mathcal C})\}(U_{{\bm k}}({\mathcal C}){\bm \psi}_{n{\bm k}}^{*})\notag\\
  &=\zeta_{\mathcal C}{\bm \psi}_{n{\bm k}}^{\dagger}j^{i}_{\bm k}(U_{{\bm k}}({\mathcal C}){\bm \psi}_{n{\bm k}}^{*}).
\end{align}
Thus, we conclude that
\begin{align}
  {\bm \psi}_{n{\bm k}}^{\dagger}j^{i}_{\bm k}(U_{{\bm k}}({\mathcal C}){\bm \psi}_{n{\bm k}}^{*})=\zeta_{\mathcal C}{\bm \psi}_{n{\bm k}}^{\dagger}j^{i}_{\bm k}(U_{{\bm k}}({\mathcal C}){\bm \psi}_{n{\bm k}}^{*}).
\end{align}
and when $\zeta_{\mathcal C}=-1$, optical excitations are suppressed.

Next, we extend the above discussion to the edge.
Here, we assume, as in Sec.~\ref{sec:edge}, that the Hamiltonian can be separated into $H_{\bm k} = h^x(k_x) + h^y(k_y)$.
The edge current and symmetry are given by
\begin{align}
  j^{\rm edge}_{k_x}&=\chi^{\dagger}j^x_{k_x}\chi,\\
  U^{\rm edge}_{k_x}({\mathcal C})&=\chi^{\dagger}U_{k_x}({\mathcal C})\chi^*.
\end{align}
By considering the transformation of $\chi$ under ${\mathcal C}$, as done in Sec.~\ref{sec:edge} and Appendix~\ref{app:edge_symmetry}, we find
\begin{align}
  U_{k_x}({\mathcal C})\chi^{\ast}=\chi W_{k_x}({\mathcal C}),\\
  \chi^{\top}U_{k_x}^{\dagger}({\mathcal C})=W_{k_x}^{\dagger}({\mathcal C})\chi^{\dagger}.
\end{align}
and using unitarity, we also have
\begin{align}
  \chi^{\ast}W^{\dagger}_{k_x}({\mathcal C})=U^{\dagger}_{k_x}({\mathcal C})\chi,\\
  W_{k_x}({\mathcal C})\chi^{\top}=\chi^{\dagger}U_{k_x}({\mathcal C}).
\end{align}
Considering the matrix element of the edge current operator between the edge states ${\bm \psi}_{nk_x}$ and $U^{\rm edge}_{\bm k}({\mathcal C}){\bm \psi}_{nk_x}$, we obtain
\begin{align}
  &\quad {\bm \psi}_{nk_x}^{\dagger}j_{k_x}^{\rm edge}(U_{k_x}^{\rm edge}({\mathcal C}){\bm \psi}_{nk_x}^*)\notag\\
  &=\zeta_{{\mathcal C}}{\bm \psi}_{nk_x}^{\dagger}[U_{k_x}^{\rm edge}({\mathcal C})(j_{k_x}^{\rm edge})^*\{U_{k_x}^{\rm edge}({\mathcal C})\}^{\dagger}](U_{k_x}^{\rm edge}({\mathcal C}){\bm \psi}_{nk_x}^*)
\end{align}
and since
\begin{align}
  &\quad U_{k_x}^{\rm edge}({\mathcal C})(j_{k_x}^{\rm edge})^*\{U_{k_x}^{\rm edge}({\mathcal C})\}^{\dagger}\notag\\
  &=\chi^{\dagger} U_{k_x}({\mathcal C})\chi^*\chi^{\top} j^{x*}_{k_x}\chi^* \chi^{\top}U^{\dagger}_{k_x}({\mathcal C})\chi\notag\\
  &=\chi^{\dagger}\chi W_{k_x}({\mathcal C})\chi^{\top} j^{x*}_{k_x}\chi^* W^{\dagger}_{k_x}({\mathcal C})\chi^{\dagger}\chi\notag\\
  &=\chi^{\dagger} U_{k_x}({\mathcal C})j^{x*}_{k_x}U^{\dagger}_{k_x}({\mathcal C})\chi\notag\\
  &=\chi^{\dagger}j^{x}_{k_x}\chi\notag\\
  &=j^{\rm edge}_{k_x},
\end{align}
we find that
\begin{align}
  {\bm \psi}_{nk_x}^{\dagger}j_{k_x}^{\rm edge}(U_{k_x}^{\rm edge}({\mathcal C}){\bm \psi}_{nk_x}^*)=\zeta_{{\mathcal C}}{\bm \psi}_{nk_x}^{\dagger}j^{\rm edge}_{k_x}(U_{k_x}^{\rm edge}({\mathcal C}){\bm \psi}_{nk_x}^*).
\end{align}
Therefore, we conclude that, similar to the bulk case, when $\zeta_{\mathcal C}=-1$, optical excitations between the edge states ${\bm \psi}_{nk_x}$ and $U^{\rm edge}_{\bm k}({\mathcal C}){\bm \psi}^*_{nk_x}$ are suppressed.
}
\bibliography{ref}

\clearpage
\onecolumngrid
\begin{center}
\large
\textbf{Supplementary Material for ``Optical response of edge modes in time-reversal symmetric topological superconductors''}
\end{center}
\setcounter{section}{0}
\setcounter{equation}{0}
\setcounter{figure}{0}
\setcounter{table}{0}
\renewcommand{\thesection}{S\arabic{section}}
\renewcommand{\theequation}{S\arabic{equation}}
\renewcommand{\thefigure}{S\arabic{figure}}
\renewcommand{\thetable}{S\arabic{table}}

\addtocontents{toc}{\protect\setcounter{tocdepth}{0}}
\section{Detailed calculations of optical conductivity}
\label{app:detail}
We provide the detailed calculations carried out in Sec.III.
We note that $\tau_i, s_i, \rho_i~(i=0,x,y,z)$ are the Pauli matrices corresponding to the Nambu space, spin space, and sublattice, respectively.

\subsection{Two dimensional ${\mathbb Z}_2$ topological insulator}
\label{app:qsh}
First, we discuss the perturbations applied to the two-dimensional ${\mathbb Z}_2$ topological insulator treated in Sec.~III\thinspace A.
The Hamiltonian matrix $H_{\bm k}$ without perturbations is given by Eq.~(45)).
The only symmetry imposed on this system is time-reversal symmetry $T$, which is represented as $U(T)=\ii \sigma_y\otimes\rho_0$.
Thus, it is possible to add a $4\times4$ matrix $M_{\bm k}$ as a perturbation that satisfies
\begin{align}
  U(T)M_{\bm k}^*=M_{-{\bm k}}U(T).
\end{align}
Such a matrix $M_{\bm k}$ can be expressed as
\begin{align}
  M_{\bm k}=\sum_{i=1}^{4}a_i({\bm k})\Gamma^{\rm even}_i+\sum_{j=1}^{6}b_j({\bm k})\Gamma^{\rm odd}_j,
\end{align}
where $a_i({\bm k})$ and $b_j({\bm k})$ are real functions that are even and odd under inversion of ${\bm k}$, respectively.
The corresponding $4\times4$ matrices are given by
\begin{align}
  \Gamma^{\rm even}_1=\sigma_0\otimes\rho_x,\quad
  \Gamma^{\rm even}_2=\sigma_x\otimes\rho_y,\quad
  \Gamma^{\rm even}_3=\sigma_y\otimes\rho_y,\quad
  \Gamma^{\rm even}_4=\sigma_z\otimes\rho_y,
\end{align}
and
\begin{align}
  &\Gamma^{\rm odd}_1=\sigma_x\otimes\rho_0,\quad
  \Gamma^{\rm odd}_2=\sigma_x\otimes\rho_x,\quad
  \Gamma^{\rm odd}_3=\sigma_x\otimes\rho_z,\quad\notag\\
  &\Gamma^{\rm odd}_4=\sigma_y\otimes\rho_0,\quad
  \Gamma^{\rm odd}_5=\sigma_y\otimes\rho_x,\quad
  \Gamma^{\rm odd}_6=\sigma_y\otimes\rho_z.
\end{align}
In the numerical calculations, we added random perturbations due to nearest-neighbor hopping as follows:
\begin{align}
  a_1({\bm k})&=0.0173498\cos k_x + 0.0276193 \cos k_y, & a_2({\bm k})&=0.0418307\cos k_x + 0.0286269 \cos k_y,\notag\\
  a_3({\bm k})&=0.0168843\cos k_x + 0.0289000 \cos k_y, & a_4({\bm k})&=0.0440222\cos k_x + 0.0287675 \cos k_y,\notag\\
  b_1({\bm k})&=0.0247570\sin k_x + 0.0369106 \sin k_y, & b_2({\bm k})&=0.0461477\sin k_x + 0.0302714 \sin k_y,\notag\\
  b_3({\bm k})&=0.0133201\sin k_x + 0.0300183 \sin k_y, & b_4({\bm k})&=0.0441472\sin k_x + 0.0296096 \sin k_y,\notag\\
  b_5({\bm k})&=0.0193532\sin k_x + 0.0285191 \sin k_y, & b_6({\bm k})&=0.0413659\sin k_x + 0.0360463 \sin k_y.
\end{align}

Next, we discuss the effective edge theory of the two dimensional ${\mathbb Z}_2$ topological insulator considered in Sec.~III\thinspace A.
Let the fermionic operator that constitutes the edge modes of the two dimensional ${\mathbb Z}_2$ topological insulator be denoted as $\hat{\bm f}^{}_{k}=(\hat{f}^{}_{\uparrow k}, \hat{f}^{}_{\downarrow k})^{\top}$.
In this case, the Hamiltonian can be written as
\begin{align}
  \label{eq:qsh_edge_Hamiltonian_app}
  \hat{H}^{\rm edge}&=\sum_{k_x}\hat{\bm f}^{\dagger}_{k_x}H^{\rm edge}_{k_x}\hat{\bm f}^{}_{k_x}.
\end{align}
Since the edge modes have time-reversal symmetry $T$, the $2\times 2$ matrix $H_{k_x}^{\rm edge}$ must satisfy
\begin{align}
  U^{\rm edge}(T)(H_{k_x}^{\rm edge})^*&=H_{-k_x}^{\rm edge}U^{\rm edge}(T),\\
  U^{\rm edge}(T)&=\ii s_y.
\end{align}
The matrices allowed under this symmetry constraint are given by
\begin{align}
  H_{k_x}^{\rm edge}=E_0(k_x)s_0+{\bm a}(k_x)\cdot{\bm s},
\end{align}
where $E_0(k_x)$ and ${\bm a}(k_x)=(a_x(k_x),a_y(k_x),a_z(k_x))$ are real functions that are even and odd with respect to $k_x$, respectively.
Thus, the Hamiltonian up to the first order in $k_x$ is
\begin{align}
  H_{k_x}^{\rm edge}=E_0 s_0+({\bm a}\cdot{\bm s})k_x =E_0 s_0+k_x
  \begin{pmatrix}
    a_z & a_x-\ii a_y\\
    a_x+\ii a_y & -a_z
  \end{pmatrix}.
\end{align}
Since $E_0$ represents an energy shift, we can set $E_0=0$, resulting in the Hamiltonian:
\begin{gather}
  \label{eq:qsh_edge_Hamiltonian_matrix}
  H^{\rm edge}_{k_x}=k_x
  \begin{pmatrix}
    a_z & a_x-ia_y\\
    a_x+ia_y & -a_z
  \end{pmatrix}.
\end{gather}
The eigenenergies and eigenstates of this Hamiltonian are given by
\begin{align}
  \epsilon_{1,k_x}&=ak_x,\qquad u_{1,k_x}=\frac{1}{\sqrt{2}\sqrt{a^2-aa_z}}\left(a_z-a,\ a_x+\ii a_y\right)^{\top},\\
  \epsilon_{2,k_x}&=-ak_x,\qquad u_{2,k_x}=\frac{1}{\sqrt{2}\sqrt{a^2+aa_z}}\left(a_z+a,\ a_x+\ii a_y\right)^{\top},
\end{align}
where $a=\sqrt{a_x^2+a_y^2+a_z^2}$.
Since the edge Hamiltonian~\eqref{eq:qsh_edge_Hamiltonian_app} possesses $U(1)$ symmetry, the current operator is obtained from the Hamiltonian matrix~\eqref{eq:qsh_edge_Hamiltonian_matrix} as
\begin{align}
  \hat{j}^{\rm edge}_{q}&=\sum_{k_x}\hat{\bm f}^{\dagger}_{k_x+q}j^{\rm edge}_{k_x,q}\hat{\bm f}^{}_{k_x},\\
  \label{eq:qsh_edge_current_matrix}
  j_{k_x,q}^{\rm edge}&=\left(1-\frac{\ii q}{2}\right)
  \begin{pmatrix}
    a_z & a_x-\ii a_y\\
    a_x+\ii a_y & -a_z
  \end{pmatrix}.
\end{align}

Then, we compute the momentum dependent optical conductivity from this effective edge theory.
The Hamiltonian~\eqref{eq:qsh_edge_Hamiltonian_matrix} and current operator~\eqref{eq:qsh_edge_current_matrix} are diagonalized by $U_{k_x}=(u_{1,k_x},u_{2,k_x})$ as
\begin{align}
  &U_{k_x}^{\dagger}H^{\rm edge}_{k_x}U_{k_x}=k_x
  \begin{pmatrix}
    a & 0\\
    0 & -a
  \end{pmatrix},\\
  &j^{\psi}_{k,q}=U_{k_x+q}^{\dagger}j^{\rm edge}_{k_x,q}U_{k_x}=\left(1-\frac{\ii q}{2}\right)
  \begin{pmatrix}
    a & 0\\
    0 & -a
  \end{pmatrix}.
\end{align}
Since the Hamiltonian and current operator are diagonalized simultaneously, only intra-band optical excitations occur.
Thus, it is evident that the optical conductivity vanishes $\sigma(\omega,q=0)=0$~[30].
For the momentum dependent optical conductivity at $T=0$, we have
\begin{align}
  &{\rm Re}[\sigma^{\rm edge}(\omega,q)]_{T=0}\notag\\
  &=\frac{1}{\omega}\sum_{n,m}\int_{-k_c}^{k_c} dk_x[f(\epsilon_{mk_x})-f(\epsilon_{nk_x+q})]|j_{m,n}(k_x,q)|^2\delta(\omega-\epsilon_{nk_x+q}+\epsilon_{mk_x})\notag\\
  &=\frac{a^2}{\omega}\left(1+\frac{q^2}{4}\right)\int_{-k_c}^{k_c} dk_x\left[\left\{f(a k_x)-f(a (k_x+q))\right\}\delta(\omega-a q)+\left\{f(-ak_x)-f(-a(k_x+q))\delta(\omega+a q)\right\}\right]\notag\\
  &=\left(1+\frac{q^2}{4}\right)\left\{\delta\left(q-\frac{\omega}{a}\right)+\delta\left(q+\frac{\omega}{a}\right)\right\}
\end{align}
in the low energy region $0 < \omega < ak_c$, where a cutoff at $k_c$ is considered.
Using this momentum dependent optical conductivity and the relation (19), we can transform it into the spatially-resolved optical conductivity. In the low energy region where $\omega\ll a$, the spatially-resolved optical conductivity becomes
\begin{align}
  {\rm Re}[\tilde{\sigma}^{\rm edge}(\omega,R_x)]&=\frac{1}{2\pi X^2}\sum_{n,n'=0}^{X-1}\int_{-\pi}^{\pi} \frac{dq}{2\pi} \cos\left\{q\cdot(n-n')\right\}{\rm Re}[\sigma(\omega,q)]\notag\\
  &=\frac{1}{2\pi^2 X^2}\sum_{n,n'=0}^{X-1}\left(1+\frac{\omega^2}{4a^2}\right)\cos\left\{\frac{\omega}{a}\cdot(n-n')\right\}\notag\\
  &\sim \frac{1}{2\pi^2 X^2}\sum_{n,n'=0}^{X-1}\notag\\
  &=\frac{1}{2\pi^2}.
\end{align}
This result shows that in the low energy region $\omega\ll a$, the spatially-resolved optical conductivity remains constant.

\subsection{Strong topological superconductor}
\label{app:helical}
We show the detailed calculation of the momentum dependent optical conductivity of edge modes in a strong topological superconductor discussed in Sec.~III\ B.
The Hamiltonian of a strong topological superconductor is given by
\begin{gather}
  \hat{H}=\frac{1}{2}\sum_{\bm k}\hat{\bm c}^{\dagger}_{\bm k}
  \begin{pmatrix}
    h_{\bm k} & \Delta_{\bm k}\\
    \Delta_{\bm k}^{\dagger} & -h_{-{\bm k}}^{\top}
  \end{pmatrix}
  \hat{\bm c}^{}_{\bm k},\\
  \hat{\bm c}^{}_{\bm k}=\left(\hat{c}^{}_{{\bm k},\uparrow},\hat{c}^{}_{{\bm k},\downarrow}, \hat{c}^{\dagger}_{-{\bm k},\uparrow},\hat{c}^{\dagger}_{-{\bm k},\downarrow}\right)^{\top},
\end{gather}
where
\begin{gather}
  h_{\bm k}=
  \begin{pmatrix}
    \mu-t\cos k_x-t\cos k_y & -\ii \nu\sin k_x\\
    \ii \nu\sin k_x & \mu-t\cos k_x-t\cos k_y
  \end{pmatrix}
  ,
  \\
  \Delta_{\bm k}=
  \begin{pmatrix}
    \Delta\sin k_x +\ii\Delta\sin k_y & 0\\
    0 & -\Delta\sin k_x +\ii\Delta\sin k_y
  \end{pmatrix}
  .
\end{gather}
The equation for the projection basis~(24) becomes
\begin{align}
  \{(\mu-2t)s_z\otimes \tau_0- s_y\otimes\tau_0 \Delta (-\ii \partial_{y})\}\varphi_i(y){\bm \chi}_i={\bm 0}.
\end{align}
Multiplying both sides by $s_z\otimes\tau_0$ from the left and rearranging, we get
\begin{align}
  \label{eq:y_zero_mode_helical}
  \frac{2t-\mu}{\Delta}\varphi_i(y){\bm \chi}_i=(\partial_y \varphi(y))s_x\otimes \tau_0 {\bm \chi}_i.
\end{align}
Considering the solutions to this equation, ${\bm \chi}_i$ must be an eigenvector of $s_x\otimes\tau_0$, satisfying
\begin{align}
  s_x\otimes\tau_0 {\bm \chi}_i=\lambda_{i}{\bm \chi}_i,
\end{align}
where $\lambda_i=\pm1$.
For such ${\bm \chi}_i$, Eq.~\eqref{eq:y_zero_mode_helical} becomes
\begin{align}
  \frac{2t-\mu}{\Delta}\varphi_i(y)=\lambda_i\partial_y \varphi(y),
\end{align}
and thus, $\varphi_i(y)$ takes the form
\begin{align}
  \varphi_i(y)\propto e^{\lambda_i\frac{2t-\mu}{\Delta}y}.
\end{align}
Depending on the sign of $\lambda_i\frac{2t-\mu}{\Delta}$, it decays exponentially in the $y$ direction.
If ${\bm \chi}_i$ is associated with $\lambda_i = +1$ for the edge we are considering,, the projection is given by
\begin{align}
  \chi=\frac{1}{\sqrt{2}}
  \begin{pmatrix}
    1 & 0\\
    0 & 1\\
    1 & 0\\
    0 & 1
  \end{pmatrix}.
\end{align}
Using this projection, the Hamiltonian and current operator are given by
\begin{align}
  &\hat{H}^{\rm edge}=\frac{1}{2}\sum_{k_x}\hat{\bm \gamma}^{\dagger}_{k_x}
  \begin{pmatrix}
    \Delta\sin k_x & 0\\
    0 & -\Delta\sin k_x
  \end{pmatrix}
  \hat{\bm \gamma}_{k_x},\\
  &\hat{j}^{\rm edge}_{q}=\frac{1}{2}\sum_{k_x}\hat{\bm \gamma}^{\dagger}_{k_x+q}
  \begin{pmatrix}
    t e^{\ii\frac{q}{2}}\sin(k_x+\tfrac{q}{2}) & -\ii \nu e^{-\ii\frac{q}{2}}\cos(k_x+\tfrac{q}{2})\\
    \ii \nu e^{-\ii\frac{q}{2}}\cos(k_x+\tfrac{q}{2}) & t e^{\ii\frac{q}{2}}\sin(k_x+\tfrac{q}{2})
  \end{pmatrix}
  \hat{\bm \gamma}_{k_x}.
\end{align}
To avoid the fermion doubling problem, we introduce a momentum cutoff $k_c$ and linearize with respect to $k_x$.
The effective edge theory then becomes
\begin{align}
  \label{eq:helical_edge_Hamiltonian}
  &\hat{H}^{\rm edge}=\frac{1}{2}\sum_{-k_c\leq k_x\leq k_c}\hat{\bm \gamma}^{\dagger}_{k_x}
  \begin{pmatrix}
    \Delta k_x & 0\\
    0 & -\Delta k_x
  \end{pmatrix}
  \hat{\bm \gamma}_{k_x},\\
  &\hat{j}^{\rm edge}_{q}=\frac{1}{2}\sum_{k_x}\Theta(-k_c\leq k_x\leq k_c)\Theta(-k_c\leq k_x+q\leq k_c)\hat{\bm \gamma}^{\dagger}_{k_x+q}
  \begin{pmatrix}
    t(k_x+\tfrac{q}{2}) & -\ii\nu(1-\ii\frac{q}{2})\\
    \ii\nu(1-\ii\frac{q}{2}) & t(k_x+\tfrac{q}{2})
  \end{pmatrix}
  \hat{\bm \gamma}_{k_x}.
\end{align}

Next, we compute the momentum-dependent optical conductivity using the effective edge theory.
Since the Hamiltonian~\eqref{eq:helical_edge_Hamiltonian} is already diagonalized, the eigenvalues and eigenstates are given by
\begin{align}
  \epsilon_{1k_x}=\Delta k_x,\quad {\bm u}_{1k_x} = (1,0)^{\top},\\
  \epsilon_{2k_x}=-\Delta k_x,\quad {\bm u}_{2k_x} = (0,1)^{\top}.
\end{align}
Thus, the matrix elements of the edge current operator in the eigenstate basis $j^{\psi}_{k_x,q}$ are
\begin{gather}
  j^{\psi}_{k_x,q}=
  \begin{pmatrix}
    t(k_x+\tfrac{q}{2}) & -\ii\nu(1-\ii\frac{q}{2})\\
    \ii\nu(1-\ii\frac{q}{2}) & -t(k_x+\tfrac{q}{2})
  \end{pmatrix}.
\end{gather}
Using these and expression~(33), the momentum dependent optical conductivity is calculated as
\begin{align}
  &{\rm Re}[\sigma^{\rm edge}(\omega,q)]\notag\\
  &=\frac{1}{4\omega}\sum_{n,m}\int_{-\pi}^{\pi} dk\{f(\epsilon_{nk_x})-f(\epsilon_{mk_x+q})\}\left|[j_{k_x,q}^{\psi}]_{nm}\right|^2\delta(\omega-\epsilon_{mk_x+q}+\epsilon_{nk_x})\notag\\
  &=\frac{t^2}{4\omega}\int_{I(q)} dk_x \left(k_x+\frac{q}{2}\right)^2 \left[\left\{f(\Delta k_x)-f(\Delta (k_x+q))\right\}\delta(\omega-\Delta q)+\left\{f(-\Delta k_x)-f(-\Delta (k_x+q))\right\}\delta(\omega+\Delta q)\right]\notag\\
  &\quad +\frac{\nu^2}{4\omega}\left(1+\frac{q^2}{4}\right)\int_{I(q)} dk_x\notag\\
  &\qquad \times\left[\left\{f(-\Delta k_x)-f(\Delta (k_x+q))\right\}\delta(\omega-\Delta(2k_x+q))+\left\{f(\Delta k_x)-f(-\Delta (k_x+q))\right\}\delta(\omega+\Delta(2k_x+q))\right],
\end{align}
where $I(q)=[-k_c,k_c]\cap[-k_c-q,k_c-q]$.
At $T=0$, the result becomes
\begin{align}
  &{\rm Re}[\sigma^{\rm edge}(\omega,q)]_{T=0}={\rm Re}[\sigma^{\rm intra}(\omega,q)]_{T=0}+{\rm Re}[\sigma^{\rm inter}(\omega,q)]_{T=0},\\
  &{\rm Re}[\sigma^{\rm intra}(\omega,q)]_{T=0}=\frac{t^2\omega^2}{48\Delta^3}\left\{\delta(\omega-\Delta q)+\delta(\omega+\Delta q)\right\}
  \times
  \begin{dcases}
    1
		\quad(0\leq\omega\leq\Delta k_c)\\
    \left(\frac{2\Delta k_c}{\omega}-1\right)^3
    \quad(\Delta k_c\leq\omega\leq2\Delta k_c)\\
    0\qquad(\omega\leq 2\Delta k_c)
  \end{dcases},\\
  &{\rm Re}[\sigma^{\rm inter}(\omega,q)]_{T=0}=\frac{\nu^2}{4\Delta\omega}\left(1+\frac{q^2}{4}\right)
  \times
  \begin{dcases}
		\Theta\left(-\frac{\omega}{\Delta}\leq q \leq \frac{\omega}{\Delta}\right)
		\quad(0\leq\omega\leq\Delta k_c)\\
		\Theta\left(-2k_c+\frac{\omega}{\Delta}\leq q \leq 2k_c-\frac{\omega}{\Delta}\right)\quad(\Delta k_c\leq\omega\leq2\Delta k_c)\\
    0\qquad(\omega\leq 2\Delta k_c)
  \end{dcases}.
\end{align}
Here, $\sigma^{\rm intra}(\omega,q)$ and $\sigma^{\rm inter}(\omega,q)$ represent contributions from intra-band and inter-band excitations, respectively.
We find that there is only a contribution from inter-band excitations in the optical conductivity.
Furthermore, using relation~(19), we can derive the spatially-resolved optical conductivity from these results.

\subsection{Crystalline superconductor in layer group $pmaa$}
\label{app:pmaa}
The Hamiltonian of the topological crystalline superconductor in the layer group $pmaa$ discussed in Sec.~III\thinspace C is given by
\begin{gather}
  \hat{H} = \frac{1}{2} \sum_{\bm{k}} \hat{\bm{c}}_{\bm{k}}^{\dagger}
  \begin{pmatrix}
    h_{\bm{k}} & \Delta_{\bm{k}}^{} \\
    \Delta_{\bm{k}}^{\dagger} & -h_{-{\bm{k}}}^{\top}
  \end{pmatrix}
  \hat{\bm{c}}_{\bm{k}}^{},
  \\
  \hat{\bm{c}}_{\bm{k}} = 
  \left( 
    \hat{c}_{{\bm{k}}, \uparrow, A}, \hat{c}_{{\bm{k}}, \downarrow, A}, 
    \hat{c}_{{\bm{k}}, \uparrow, B}, \hat{c}_{{\bm{k}}, \downarrow, B}, 
    \hat{c}^{\dagger}_{-{\bm{k}}, \uparrow, A}, \hat{c}^{\dagger}_{-{\bm{k}}, \downarrow, A}, 
    \hat{c}^{\dagger}_{-{\bm{k}}, \uparrow, B}, \hat{c}^{\dagger}_{-{\bm{k}}, \downarrow, B} 
  \right)^{\top},
\end{gather}
where the normal-phase Hamiltonian and the superconducting order parameter are given by
\begin{gather}
  h_{\bm{k}} =
  \resizebox{0.9\columnwidth}{!}{$
  \begin{pmatrix}
    -t_x (\cos k_x - 1) - t_y \cos k_y & t_1 \sin k_x & t_2 (1 + e^{-\ii k_x}) & 0 \\
    t_1 \sin k_x & -t_x (\cos k_x - 1) - t_y \cos k_y & 0 & t_2 (1 + e^{-\ii k_x}) \\
    t_2 (1 + e^{\ii k_x}) & 0 & -t_x (\cos k_x - 1) - t_y \cos k_y & -t_1 \sin k_x \\
    0 & t_2 (1 + e^{\ii k_x}) & -t_1 \sin k_x & -t_x (\cos k_x - 1) - t_y \cos k_y
  \end{pmatrix}
  ,
  $}
  \\
  \Delta_{\bm{k}} =
  \begin{pmatrix}
    m_1 \sin k_x + \ii\Delta \sin k_y & -m_3 - m_4 (\cos k_x - 1) & 0 & m_2 (1 + e^{-\ii k_x}) \\
    m_3 + m_4 (\cos k_x - 1) & -m_1 \sin k_x + \ii \Delta \sin k_y & -m_2 (1 + e^{-\ii k_x}) & 0 \\
    0 & m_2 (1 + e^{\ii k_x}) & -m_1 \sin k_x - \ii \Delta \sin k_y & -m_3 - m_4 (\cos k_x - 1) \\
    -m_2 (1 + e^{\ii k_x}) & 0 & m_3 + m_4 (\cos k_x - 1) & m_1 \sin k_x - \ii\Delta \sin k_y
  \end{pmatrix}
  .s
\end{gather}

First, we consider the EAZ class of the edge to determine whether optical conductivity exists in the low energy regime.
As discussed in Sec.~III\thinspace C, the states of this system are classified into eigenvalue sectors based on the glide symmetry $G_z$ with eigenvalue $\xi_{\bm{k}}(G_z) = \pm \ii e^{-\ii k_x / 2}$.
We then examine whether the symmetry ${\mathcal T} = M_x T$ and ${\mathcal C} = M_x C$ are closed within each eigenvalue sector.
For ${\mathcal T}$, we find that $(T_x^{-1} G_z)(M_x T) = (M_x T)(G_z)$ holds, where $T_x$ is the translation in the $x$-direction,.
In the projective representation, the projective factor is $z_{G_z, M_x T} = -z_{M_x T, G_z} = +1$.
Using the relationship in Eq.~(B4), the eigenvalue $\xi'_{\bm{k}}(G_z)$ of the transformed state $U_{\bm{k}}({\mathcal T}) {\bm \psi}^*_{n{\bm{k}}}$ becomes
\begin{align}
  \xi'_{\bm{k}}(T_x^{-1} G_z) &= e^{\ii k_x} \xi'_{\bm{k}}(G_z) \notag\\
  &= \frac{z_{T_x^{-1} G_z, M_x T}}{z_{M_x T, G_z}} (\xi_{\bm{k}}(G_z))^* \notag\\
  &= -(\mp \ii e^{\ii k_x / 2}) \notag\\
  &= \pm \ii e^{\ii k_x / 2}.
\end{align}
This shows that $\xi'_{\bm{k}}(G_z) = \xi_{\bm{k}}(G_z)$, meaning that ${\mathcal T}$ is closed within each $G_z$ eigenvalue sector.
Similarly, for ${\mathcal C}$, we have $(T_x^{-1} G_z)(M_x C) = (M_x C)(G_z)$ and $z_{G_z, M_x C} = -z_{M_x C, G_z} = +1$.
Thus, the eigenvalue $\xi'_{\bm{k}}(G_z)$ of the transformed state $U_{\bm{k}}({\mathcal C}) {\bm \psi}^*_{n{\bm{k}}}$ becomes
\begin{align}
  \xi'_{\bm{k}}(T_x^{-1} G_z) &= e^{\ii k_x} \xi'_{\bm{k}}(G_z) \notag\\
  &= \frac{z_{T_x^{-1} G_z, M_x C}}{z_{M_x C, G_z}} (\xi_{\bm{k}}(G_z))^* \notag\\
  &= -(\mp \ii e^{\ii k_x / 2}) \notag\\
  &= \pm \ii e^{\ii k_x / 2}.
\end{align}
Therefore, ${\mathcal C}$ is also closed within each $G_z$ eigenvalue sector.
Since both ${\mathcal T}$ and ${\mathcal C}$ are closed within each eigenvalue sector, the chiral symmetry $S$ is also closed within each eigenvalue sector.
As $z_{M_x T, M_x T} = -z_{M_x C, M_x C} = +1$, the edge belongs to the class CI in the EAZ class.
From Table~I, we conclude that there is no optical conductivity in the low energy regime.

Next, we consider an effective edge theory to calculate the momentum-dependent optical conductivity.
When $t_i$ and $m_i$ are sufficiently small, the relation satisfied by the edge modes, as derived from Eq.~(24), is
\begin{align}
  [-t_y \tau_z \otimes \rho_0 \otimes \sigma_0 - \Delta \tau_y \otimes \rho_z \otimes \sigma_0 (-\ii \partial_y)] \varphi_i(y) {\bm \chi}_i = {\bm 0}.
\end{align}s
Rearranging this equation, we obtain
\begin{align}
  \label{eq:y_zero_mode_pmaa}
  \frac{t_y}{\Delta} \varphi_i(y) {\bm \chi}_i = (\partial_y \varphi_i(y)) \tau_x \otimes \rho_z \otimes \sigma_0 {\bm \chi}_i.
\end{align}
Thus, ${\bm \chi}_i$ is an eigenvector of $\tau_x \otimes \rho_z \otimes \sigma_0$, corresponding to the eigenvalue $\lambda_i = \pm 1$, satisfying
\begin{align}
  \tau_x \otimes \rho_z \otimes \sigma_0 {\bm \chi}_i = \lambda_i {\bm \chi}_i.
\end{align}
For such ${\bm \chi}_i$, the equation becomes
\begin{align}
  \frac{t_y}{\Delta} \varphi_i(y) = \lambda_i(\partial_y \varphi_i(y)).
\end{align}
Thus, $\varphi_i(y)$ has the form
\begin{align}
  \varphi_i(y) \propto e^{\lambda_i \frac{t_y}{\Delta} y}.
\end{align}
If the basis corresponding to $\lambda_i = +1$ is chosen for the edge of interest, the projection is given by
\begin{align}
  \chi = \frac{1}{\sqrt{2}}
  \begin{pmatrix}
    1 & 0 & 0 & 0 \\
    0 & 1 & 0 & 0 \\
    0 & 0 & -\ii & 0 \\
    0 & 0 & 0 & -\ii \\
    1 & 0 & 0 & 0 \\
    0 & 1 & 0 & 0 \\
    0 & 0 & \ii & 0 \\
    0 & 0 & 0 & \ii
  \end{pmatrix}.
\end{align}
Using this projection, the Hamiltonian and the current operator are given by
\begin{align}
  \label{eq:effective_Hamiltonian_pmaa}
  &\hat{H}^{\rm eff} = \frac{1}{2} \sum_{k_x} \hat{\bm \gamma}_{k_x}^{\dagger}
  \begin{pmatrix}
    m_1 \sin k_x & t_1 \sin k_x & -\ii t_2 (1 + e^{-\ii k_x}) & \ii m_2 (1 + e^{-\ii k_x}) \\
    t_1 \sin k_x & -m_1 \sin k_x & -\ii m_2 (1 + e^{-\ii k_x}) & -\ii t_2 (1 + e^{-\ii k_x}) \\
    \ii t_2 (1 + e^{\ii k_x}) & \ii m_2 (1 + e^{\ii k_x}) & m_1 \sin k_x & -t_1 \sin k_x \\
    -\ii m_2 (1 + e^{\ii k_x}) & \ii t_2 (1 + e^{\ii k_x}) & -t_1 \sin k_x & -m_1 \sin k_x
  \end{pmatrix}
  \hat{\bm \gamma}_{k_x},\\
  &\hat{j}^{\rm edge}_q = \frac{1}{2} \sum_{k} \{-t_x e^{-\ii \frac{q}{2}} \sin (k_x + \tfrac{q}{2})\} \hat{\bm \gamma}^{\dagger}_{k_x}
  \begin{pmatrix}
    1 & 0 & 0 & 0 \\
    0 & 1 & 0 & 0 \\
    0 & 0 & 1 & 0 \\
    0 & 0 & 0 & 1
  \end{pmatrix}
  \hat{\bm \gamma}_{k_x}.\label{j_gamma_pmaa}
\end{align}

We then calculate the momentum dependent optical conductivity using the derived effective edge theory.
First, the eigenvalues of the Hamiltonian~\eqref{eq:effective_Hamiltonian_pmaa} are given by
\begin{gather}
  \epsilon_{\pm 1k_x} = \pm 2 \cos \frac{k_x}{2} \sqrt{\{a_{\pm}(k_x)\}^2 + \{b_{\pm}(k_x)\}^2}, \quad
  \epsilon_{\pm 2k_x} = \pm 2 \cos \frac{k_x}{2} \sqrt{\{a_{\mp}(k_x)\}^2 + \{b_{\mp}(k_x)\}^2},
\end{gather}
where
\begin{gather}
  a_{\pm}(k_x) \equiv m_2 \pm t_1 \sin \frac{k_x}{2}, \quad
  b_{\pm}(k_x) \equiv t_2 \pm m_1 \sin \frac{k_x}{2}.
\end{gather}
Furthermore, by introducing
\begin{align}
  c_{\pm}(k_x) \equiv b_{\pm}(k_x) - \sqrt{\{a_{\pm}(k_x)\}^2 + \{b_{\pm}(k_x)\}^2}, \quad
  d_{\pm}(k_x) \equiv b_{\pm}(k_x) + \sqrt{\{a_{\pm}(k_x)\}^2 + \{b_{\pm}(k_x)\}^2},
\end{align}
the corresponding eigenstates are given by
\begin{align}
  &{\bm u}_{\pm 1k_x} = \frac{1}{\sqrt{2} \sqrt{\{a_{\pm}(k_x)\}^2 + \{c_{\pm}(k_x)\}^2}} \left(a_{\pm}(k_x),\ -c_{\pm}(k_x),\ \pm \ii e^{\ii \frac{k_x}{2}} a_{\pm}(k_x),\ \pm \ii e^{\ii \frac{k_x}{2}} c_{\pm}(k_x)\right)^{\top}, \\
  &{\bm u}_{\pm 2k_x} = \frac{1}{\sqrt{2} \sqrt{\{a_{\mp}(k_x)\}^2 + \{d_{\mp}(k_x)\}^2}} \left(a_{\mp}(k_x),\ -d_{\mp}(k_x),\ \mp \ii e^{\ii \frac{k_x}{2}} a_{\mp}(k_x),\ \mp \ii e^{\ii \frac{k_x}{2}} d_{\mp}(k_x)\right)^{\top}.
\end{align}
The matrix elements of the current operator in this eigenstate basis $ \hat{\bm \psi}_{k} = ({\bm u}_{1k}, {\bm u}_{-1k}, {\bm u}_{2k}, {\bm u}_{-2k})\hat{\bm \gamma}_{k}$ are
\begin{align}
  &j^{\psi}_{k_x,q} = -t_x e^{-\ii \frac{3q}{4}} \sin (k_x + \tfrac{q}{2})
  \begin{pmatrix}
    j^{\psi}_{11}(k_x,q) & j^{\psi}_{12}(k_x,q) \\
    j^{\psi}_{21}(k_x,q) & j^{\psi}_{22}(k_x,q)
  \end{pmatrix},
\end{align}
where
\begin{align}
  &j^{\psi}_{11}(k_x,q)=
  \begin{pmatrix}
    \frac{\{a_{+}(k_x)a_{+}(k_x+q)+c_{+}(k_x)c_{+}(k_x+q)\}\cos\tfrac{q}{4}}{\sqrt{\{a_{+}(k_x)\}^2+\{c_{+}(k_x)\}^2}\sqrt{\{a_{+}(k_x+q)\}^2+\{c_{+}(k_x+q)\}^2}}&
    \frac{\ii\{a_{-}(k_x)a_{+}(k_x+q)+c_{-}(k_x)c_{+}(k_x+q)\}\sin\tfrac{q}{4}}{\sqrt{\{a_{-}(k_x)\}^2+\{c_{-}(k_x)\}^2}\sqrt{\{a_{+}(k_x+q)\}^2+\{c_{+}(k_x+q)\}^2}}\\
    \frac{\ii\{a_{+}(k_x)a_{-}(k_x+q)+c_{+}(k_x)c_{-}(k_x+q)\}\sin\tfrac{q}{4}}{\sqrt{\{a_{+}(k_x)\}^2+\{c_{+}(k_x)\}^2}\sqrt{\{a_{-}(k_x+q)\}^2+\{c_{-}(k_x+q)\}^2}}&
    \frac{\{a_{-}(k_x)a_{-}(k_x+q)+c_{-}(k_x)c_{-}(k_x+q)\}\cos\tfrac{q}{4}}{\sqrt{\{a_{-}(k_x)\}^2+\{c_{-}(k_x)\}^2}\sqrt{\{a_{-}(k_x+q)\}^2+\{c_{-}(k_x+q)\}^2}}
  \end{pmatrix},\\
  &j^{\psi}_{12}(k_x,q)=
  \begin{pmatrix}
    \frac{\ii\{a_{-}(k_x)a_{+}(k_x+q)+d_{-}(k_x)c_{+}(k_x+q)\}\sin\tfrac{q}{4}}{\sqrt{\{a_{-}(k_x)\}^2+\{d_{-}(k_x)\}^2}\sqrt{\{a_{+}(k_x+q)\}^2+\{c_{+}(k_x+q)\}^2}}&
    \frac{\{a_{+}(k_x)a_{+}(k_x+q)+d_{+}(k_x)c_{+}(k_x+q)\}\cos\tfrac{q}{4}}{\sqrt{\{a_{+}(k_x)\}^2+\{d_{+}(k_x)\}^2}\sqrt{\{a_{+}(k_x+q)\}^2+\{c_{+}(k_x+q)\}^2}}\\
    \frac{\{a_{-}(k_x)a_{-}(k_x+q)+d_{-}(k_x)c_{-}(k_x+q)\}\cos\tfrac{q}{4}}{\sqrt{\{a_{-}(k_x)\}^2+\{d_{-}(k_x)\}^2}\sqrt{\{a_{-}(k_x+q)\}^2+\{c_{-}(k_x+q)\}^2}}&
    \frac{\ii\{a_{+}(k_x)a_{-}(k_x+q)+d_{+}(k_x)c_{-}(k_x+q)\}\sin\tfrac{q}{4}}{\sqrt{\{a_{+}(k_x)\}^2+\{d_{+}(k_x)\}^2}\sqrt{\{a_{-}(k_x+q)\}^2+\{c_{-}(k_x+q)\}^2}}
  \end{pmatrix},\\
  &j^{\psi}_{21}(k_x,q)=
  \begin{pmatrix}
    \frac{\ii\{a_{+}(k_x)a_{-}(k_x+q)+c_{+}(k_x)d_{-}(k_x+q)\}\sin\tfrac{q}{4}}{\sqrt{\{a_{+}(k_x)\}^2+\{c_{+}(k_x)\}^2}\sqrt{\{a_{-}(k_x+q)\}^2+\{d_{-}(k_x+q)\}^2}}&
    \frac{\{a_{-}(k_x)a_{-}(k_x+q)+c_{-}(k_x)d_{-}(k_x+q)\}\cos\tfrac{q}{4}}{\sqrt{\{a_{-}(k_x)\}^2+\{c_{-}(k_x)\}^2}\sqrt{\{a_{-}(k_x+q)\}^2+\{d_{-}(k_x+q)\}^2}}\\
    \frac{\{a_{+}(k_x)a_{+}(k_x+q)+c_{+}(k_x)d_{+}(k_x+q)\}\cos\tfrac{q}{4}}{\sqrt{\{a_{+}(k_x)\}^2+\{c_{+}(k_x)\}^2}\sqrt{\{a_{+}(k_x+q)\}^2+\{d_{+}(k_x+q)\}^2}}&
    \frac{\ii\{a_{-}(k_x)a_{+}(k_x+q)+c_{-}(k_x)d_{+}(k_x+q)\}\sin\tfrac{q}{4}}{\sqrt{\{a_{-}(k_x)\}^2+\{c_{-}(k_x)\}^2}\sqrt{\{a_{+}(k_x+q)\}^2+\{d_{+}(k_x+q)\}^2}}
  \end{pmatrix},\\
  &j^{\psi}_{22}(k_x,q)=
  \begin{pmatrix}
    \frac{\{a_{-}(k_x)a_{-}(k_x+q)+d_{-}(k_x)d_{-}(k_x+q)\}\cos\tfrac{q}{4}}{\sqrt{\{a_{-}(k_x)\}^2+\{d_{-}(k_x)\}^2}\sqrt{\{a_{-}(k_x+q)\}^2+\{d_{-}(k_x+q)\}^2}}&
    \frac{\ii\{a_{+}(k_x)a_{-}(k_x+q)+d_{+}(k_x)d_{-}(k_x+q)\}\sin\tfrac{q}{4}}{\sqrt{\{a_{+}(k_x)\}^2+\{d_{+}(k_x)\}^2}\sqrt{\{a_{-}(k_x+q)\}^2+\{d_{-}(k_x+q)\}^2}}\\
    \frac{\ii\{a_{-}(k_x)a_{+}(k_x+q)+d_{-}(k_x)d_{+}(k_x+q)\}\sin\tfrac{q}{4}}{\sqrt{\{a_{-}(k_x)\}^2+\{d_{-}(k_x)\}^2}\sqrt{\{a_{+}(k_x+q)\}^2+\{d_{+}(k_x+q)\}^2}}&
    \frac{\{a_{+}(k_x)a_{+}(k_x+q)+d_{+}(k_x)d_{+}(k_x+q)\}\cos\tfrac{q}{4}}{\sqrt{\{a_{+}(k_x)\}^2+\{d_{+}(k_x)\}^2}\sqrt{\{a_{+}(k_x+q)\}^2+\{d_{+}(k_x+q)\}^2}}
  \end{pmatrix}.
\end{align}
First, when considering the optical conductivity, it follows that $\sigma^{\rm edge}(\omega,q=0)=0$ since $j^{\psi}_{k_x,q}$ is diagonal. This result is consistent with the argument based on symmetry.
Next, we examine the momentum dependent conductivity. However, $j^{\psi}_{k_x,q}$ is generally more complicated. Therefore, we limit our discussion to specific parameter settings, as described below:
\begin{description}[]
  \item[(i) $\bf{t_2=m_1=0,\ t_1\neq0,\ m_2\neq0}$] $b_{\pm}(k_x)=0$ and $c_{\pm}(k_x)=d_{\pm}(k_x)=-a_{\pm}(k_x)$\\
  \begin{align}
    &\epsilon_{\pm1k}(k_x)=\pm2\cos\frac{k_x}{2}\left(m_2\pm t_1\sin\frac{k_x}{2}\right),\quad {\bm u}_{\pm 1k_x}=\frac{1}{2}\left(1,\ 1,\ \pm \ii e^{\ii\frac{k_x}{2}},\ \mp \ii e^{\ii\frac{k_x}{2}}\right)^{\top},\\
    &\epsilon_{\pm2k}(k_x)=\pm2\cos\frac{k_x}{2}\left(m_2\mp t_1\sin\frac{k_x}{2}\right),\quad {\bm u}_{\pm 2k_x}=\frac{1}{2}\left(1,\ -1,\ \mp \ii e^{-\ii\frac{k_x}{2}},\ \mp ie^{-\ii\frac{k_x}{2}}\right)^{\top}.
  \end{align}
  \item[(ii) $\bf{t_1=m_2=0,\ t_2\neq0,\ m_1\neq0}$] $a_{\pm}(k_x)=c_{\pm}(k_x)=0$ and $d_{\pm}(k_x)=2b_{\pm}(k_x)$\\
  \begin{align}
    &\epsilon_{\pm1k}(k_x)=\pm2\cos\frac{k_x}{2}\left(t_2\pm m_1\sin\frac{k_x}{2}\right),\quad {\bm u}_{\pm 1k_x}=\frac{1}{\sqrt{2}}\left(1,\ 0,\ \pm \ii e^{\ii\frac{k_x}{2}},\ 0\right)^{\top},\\
    &\epsilon_{\pm2k_x}(k_x)=\pm2\cos\frac{k_x}{2}\left(t_2\mp m_1\sin\frac{k_x}{2}\right),\quad {\bm u}_{\pm 2k_x}=\frac{1}{\sqrt{2}}\left(0,\ 1,\ 0,\ \pm \ii e^{s\ii\frac{k_x}{2}}\right)^{\top}.
  \end{align}
  \item[(iii) $\bf{m_1=t_1, m_2=t_2}$]$b_{\pm}(k_x)=a_{\pm}(k_x)$, $c_{\pm}(k_x)=(1-\sqrt{2})a_{\pm}(k_x)$ and $d_{\pm}(k_x)=(1+\sqrt{2})a_{\pm}(k_x)$\\
  \begin{align}
    &\epsilon_{\pm1k_x}(k_x)=\pm2\sqrt{2}\cos\frac{k_x}{2}\left(t_2\pm t_1\sin\frac{k_x}{2}\right),\quad {\bm u}_{\pm 1k_x}=\frac{1}{2\sqrt{2-\sqrt{2}}}\left(1,\ \sqrt{2}-1,\ \pm \ii e^{\ii\frac{k_x}{2}},\ \mp \ii(\sqrt{2}-1)e^{\ii\frac{k_x}{2}}\right)^{\top},\\
    &\epsilon_{\pm2k_x}(k_x)=\pm2\sqrt{2}\cos\frac{k_x}{2}\left(t_2\mp t_1\sin\frac{k_x}{2}\right),\quad {\bm u}_{\pm 2k_x}=\frac{1}{2\sqrt{2+\sqrt{2}}}\left(1,\ -(\sqrt{2}+1),\ \mp \ii e^{\ii\frac{k_x}{2}},\ \mp \ii(\sqrt{2}+1)e^{\ii\frac{k_x}{2}}\right)^{\top}.
  \end{align}
\end{description}
In these parameter settings, $j^{\psi}_{k_x,q}$ is given by
\begin{align}
j^{\psi}_{k_x, q} = -t_x e^{-\ii\frac{3q}{4}} \sin \left(k_x + \tfrac{q}{2}\right)
\begin{pmatrix}
\cos \tfrac{q}{4} & \ii \sin \tfrac{q}{4} & 0 & 0\\
\ii \sin \tfrac{q}{4} & \cos \tfrac{q}{4} & 0 & 0\\
0 & 0 & \cos \tfrac{q}{4} & \ii \sin \tfrac{q}{4}\\
0 & 0 & \ii \sin \tfrac{q}{4} & \cos \tfrac{q}{4}
\end{pmatrix}
.
\end{align}
Then, the momentum dependent optical conductivity is given by
\begin{align}
    \label{eq:sigma_pmaa}
  &{\rm Re}[\sigma^{\rm edge}(\omega,q)]\notag\\
  &=\frac{1}{4\omega}\sum_{n,m}\int_{-\pi}^{\pi} dk\{f(\epsilon_{nk_x})-f(\epsilon_{mk_x+q})\}\left|[j_{k_x,q}^{\psi}]_{nm}\right|^2\delta(\omega-\epsilon_{mk_x+q}+\epsilon_{nk_x})\notag\\
  &=\frac{t_x^2}{4\omega}\cos^2\left(\frac{q}{4}\right)\int dk_x \sin^2\left(k_x+\frac{q}{2}\right)\notag\\
  &\qquad \times\left[\left\{f\left(\epsilon_{1k_x}\right)-f\left(\epsilon_{1k_x+q}\right)\right\}\delta\left(\omega-\epsilon_{1k_x+q}+\epsilon_{1k_x}\right)+\left\{f\left(\epsilon_{2k_x}\right)-f\left(\epsilon_{2k_x+q}\right)\right\}\delta\left(\omega-\epsilon_{2k_x+q}+\epsilon_{2k_x}\right)\right.\notag\\
  &\qquad \left.+\left\{f\left(\epsilon_{-1k_x}\right)-f\left(\epsilon_{-1k_x+q}\right)\right\}\delta\left(\omega-\epsilon_{-1k_x+q}+\epsilon_{-1k_x}\right)+\left\{f\left(\epsilon_{-2k_x}\right)-f\left(\epsilon_{-2k_x+q}\right)\right\}\delta\left(\omega-\epsilon_{-2k_x+q}+\epsilon_{-2k_x}\right)\right]\notag\\
  &\quad +\frac{t_x^2}{4\omega}\sin^2\left(\frac{q}{4}\right)\int dk_x \sin^2\left(k_x+\frac{q}{2}\right)\notag\\
  &\qquad \times\left[\left\{f\left(\epsilon_{-1k_x}\right)-f\left(\epsilon_{1k_x+q}\right)\right\}\delta\left(\omega-\epsilon_{1k_x+q}+\epsilon_{-1k_x}\right)+\left\{f\left(\epsilon_{-2k_x}\right)-f\left(\epsilon_{2k_x+q}\right)\right\}\delta\left(\omega-\epsilon_{2k_x+q}+\epsilon_{-2k_x}\right)\right.\notag\\
  &\qquad \left.+\left\{f\left(\epsilon_{1k}\right)-f\left(\epsilon_{-1k_x+q}\right)\right\}\delta\left(\omega-\epsilon_{-1k_x+q}+\epsilon_{1k_x}\right)+\left\{f\left(\epsilon_{2k_x}\right)-f\left(\epsilon_{-2k_x+q}\right)\right\}\delta\left(\omega-\epsilon_{-2k_x+q}+\epsilon_{2k}\right)\right].
\end{align}
Finally, using relation~(19), we obtain the spatially-resolved optical conductivity.

\subsection{Crystalline superconductor in layer group $p11a$}
\label{app:p11a}
The Hamiltonian of the topological crystalline superconductor in the layer group $p11a$ discussed in Sec.~III\thinspace D is given by
\begin{gather}
  \hat{H}=\frac{1}{2}\sum_{\bm k}\hat{\bm c}^{\dagger}_{\bm k}
  \begin{pmatrix}
    h_{\bm k} & \Delta_{\bm k}^{}\\
    \Delta_{\bm k}^{\dagger} & -h_{-{\bm k}}^{\top}
  \end{pmatrix}
  \hat{\bm c}^{}_{\bm k},\\
  \hat{\bm c}^{}_{\bm k}=\left(\hat{c}^{}_{{\bm k},\uparrow,A},\hat{c}^{}_{{\bm k},\downarrow,A},\hat{c}^{}_{{\bm k},\uparrow,B},\hat{c}^{}_{{\bm k},\downarrow,B}, \hat{c}^{\dagger}_{-{\bm k},\uparrow,A},\hat{c}^{\dagger}_{-{\bm k},\downarrow,A}, \hat{c}^{\dagger}_{-{\bm k},\uparrow,A},\hat{c}^{\dagger}_{-{\bm k},\downarrow,B}\right)^{\top},
\end{gather}
where the normal-phase Hamiltonian and superconducting order parameter are given by
\begin{gather}
  h_{{\bm k}}=
  \resizebox{0.9\columnwidth}{!}{$
  \begin{pmatrix}
    -t\cos k_y & 0 & t_1(1+e^{-\ii k_x})+\ii t_2(1-e^{-\ii k_x}) & (\ii t_3+t_4)(1+e^{-\ii k_x})\\
    0 & -t\cos k_y & (\ii t_3-t_4)(1+e^{-\ii k_x}) & t_1(1+e^{-\ii k_x})-\ii t_2(1-e^{-\ii k_x})\\
    t_1(1+e^{\ii k_x})-\ii t_2(1-e^{\ii k_x}) & -(\ii t_3+t_4)(1+e^{\ii k_x}) & -t\cos k_y & 0\\
    -(\ii t_3-t_4)(1+e^{\ii k_x}) & t_1(1+e^{\ii k_x})+\ii t_2(1-e^{\ii k_x}) & 0 & -t\cos k_y
  \end{pmatrix}
  $},\\
  \Delta_{\bm k}=
  \resizebox{0.9\columnwidth}{!}{$
  \begin{pmatrix}
    \Delta\sin k_y & 0 & (m_3+\ii m_4)(1-e^{-\ii k_x}) & m_1(1-e^{-\ii k_x})-\ii m_2(1+e^{-\ii k_x})\\
    0 & -\Delta\sin k_y & -m_1(1-e^{-\ii k_x})-\ii m_2(1+e^{-\ii k_x}) & (m_3-\ii m_4)(1-e^{-\ii k_x})\\
    -(m_3+\ii m_4)(1-e^{\ii k_x}) & m_1(1-e^{\ii k_x})+\ii m_2(1+e^{-\ii k_x}) & \Delta \sin k_y & 0\\
    -m_1(1-e^{\ii k_x})+\ii m_2(1+e^{\ii k_x}) & -(m_3-\ii m_4)(1-e^{\ii k_x}) & 0 & -\Delta \sin k_y
  \end{pmatrix}
  $}.
\end{gather}

As with the case of the layer group $pmaa$, we determine whether optical conductivity exists in the low-energy regime by analyzing the EAZ class of the edge.
The states of this system are divided into eigenvalue sectors by the glide symmetry $G_z$ with eigenvalues $\xi_{\bm k}(G_z)=\pm\ii e^{-\ii k_x/2}$ as mentioned in Sec.~III\thinspace D.
Moreover, since the edge of this system lacks crystal symmetry that transforms $k_x$ to $-k_x$, ${\mathcal T}$ and ${\mathcal C}$ are absent.
We then examine whether the chiral symmetry $S=TC$ is closed in these eigenvalue sectors.
For the chiral symmetry, $G_zS=SG_z$ and $z_{G_z,S}=-z_{S,G_z}=+1$.
Using the relationship of the projective factor~(B4), the eigenvalue $\xi'_{\bm k}(G_z)$ of the transformed state $U(T)U^*(C){\bm \psi}_{n{\bm k}}$ is given by
\begin{align}
  \xi'_{\bm k}(G_z)
  &=\frac{z_{G_z,S}}{z_{S,G_z}}\xi_{\bm k}(G_z)\notag\\
  &=-(\pm \ii e^{-\ii k_x/2})\notag\\
  &=\mp \ii e^{\ii k_x/2}
\end{align}
Thus, the chiral symmetry $S$ is not closed in the eigenvalue sectors of $G_z$ since $\xi'_{\bm k}(G_z)\neq \xi_{\bm k}(G_z)$.
Consequently, the edge of the topological crystalline superconductor in the layer group $p11a$ belongs to class A in the EAZ classification.
From Table~I, we conclude that optical conductivity is allowed in this system.
However, since the optical excitations allowed in class A do not occur between states related by ${\mathcal C}$ or chiral symmetry, the energy range where optical conductivity exists differs from the system's superconducting gap.
Therefore, while the edge modes of the current system are gapless as shown in Fig.~8, optical conductivity may not exist $\omega \sim 0$.

Then, we calculate the momentum dependent optical conductivity using the effective edge theory.
When $t_i$ and $m_i$ are sufficiently small, the equation that the edge mode basis satisfies~(24) is
\begin{align}
  [-t \tau_z\otimes\rho_0\otimes\sigma_0+\Delta\tau_x\otimes\rho_0\otimes\sigma_z(-\ii\partial_y)]\varphi_i(y){\bm \chi}_i={\bm 0}
\end{align}
Simplifying this equation yields
\begin{align}
  \label{eq:y_zero_mode_p11a}
  \frac{t}{\Delta}\varphi_i(y){\bm \chi}=(\partial_y\varphi_i(y))\tau_y\otimes\rho_0\otimes\sigma_z{\bm \chi}_i
\end{align}
Thus, ${\bm \chi}_i$ is an eigenvector of $\tau_y\otimes\rho_0\otimes\sigma_z$, and for the eigenvalue $\lambda_i=\pm1$, it satisfies
\begin{align}
  \tau_y\otimes\rho_0\otimes\sigma_z{\bm \chi}_i=\lambda_i{\bm \chi}_i
\end{align}
For such ${\bm \chi}_i$, equation~\eqref{eq:y_zero_mode_p11a} becomes
\begin{align}
  \frac{t}{\Delta}\varphi_i(y)=(\partial_y\varphi_i(y))
\end{align}
Therefore, the solution for $f_i(y)$ decays as
\begin{align}
  f_i(y)\propto e^{\lambda_i\frac{t_y}{\Delta}y}
\end{align}
For the edge of interest, if the basis corresponds to $\lambda_i=+1$, the projection is
\begin{align}
  \chi&=\frac{1}{\sqrt{2}}
  \begin{pmatrix}
    e^{-\ii\frac{\pi}{4}} & 0 & 0 & 0\\
    0 & e^{\ii\frac{\pi}{4}} & 0 & 0\\
    0 & 0 & e^{-\ii\frac{\pi}{4}} & 0\\
    0 & 0 & 0 & e^{\ii\frac{\pi}{4}}\\
    e^{\ii\frac{\pi}{4}} & 0 & 0 & 0\\
    0 & e^{-\ii\frac{\pi}{4}} & 0 & 0\\
    0 & 0 & e^{\ii\frac{\pi}{4}} & 0\\
    0 & 0 & 0 & e^{-\ii\frac{\pi}{4}}\\
  \end{pmatrix}
  ,
\end{align}
With this projection, the Hamiltonian and current operator are given by
\begin{align}
  \label{eq:edge_Hamiltonian_p11a}
  \hat{H}^{\rm edge}&=\frac{1}{2}\sum_{k_x}\hat{\bm \gamma}^{\dagger}_{k_x}
  \resizebox{0.8\columnwidth}{!}{$
  \begin{pmatrix}
    0 & 0 & -\ii(m_3-t_2)(1-e^{-\ii k_x}) & -\ii(m_2+t_4)(1+e^{-\ii k_x})\\
    0 & 0 & -\ii(m_2+t_4)(1+e^{-\ii k_x}) & \ii(m_3-t_2)(1-e^{-\ii k_x})\\
    \ii(m_3-t_2)(1-e^{\ii k_x}) & \ii(m_2+t_4)(1+e^{\ii k_x}) & 0 & 0\\
    \ii(m_2+t_4)(1+e^{\ii k_x}) & -\ii(m_3-t_2)(1-e^{\ii k_x}) & 0 & 0
  \end{pmatrix}
  $}
  \hat{\bm \gamma}^{}_{k_x},\\
  \label{eq:edge_current_p11a}
  \hat{j}^{\rm edge}_{q}&=\frac{1}{2}\sum_{k_x} \hat{\bm \gamma}^{\dagger}_{k_x+q}
  \begin{pmatrix}
    0 & 0 & \frac{\ii t_1}{2}(1-e^{-\ii (k_x+q)}) & -\frac{\ii t_3}{2}(1-e^{-\ii (k_x+q)})\\
    0 & 0 & \frac{\ii t_3}{2}(1-e^{-\ii (k_x+q)}) & \frac{\ii t_1}{2}(1-e^{-\ii (k_x+q)})\\
    -\frac{\ii t_1}{2}(1-e^{\ii k_x}) & -\frac{\ii t_3}{2}(1-e^{\ii k_x}) & 0 & 0\\
    \frac{\ii t_3}{2}(1-e^{\ii k_x}) & -\frac{\ii t_1}{2}(1-e^{\ii k_x}) & 0 & 0\\
  \end{pmatrix}
  \hat{\bm \gamma}_{k_x}.
\end{align}

We then calculate the momentum dependent optical conductivity using the effective edge theory.
The eigenvalues and eigenvectors of the Hamiltonian~\eqref{eq:edge_Hamiltonian_p11a} are given by
\begin{align}
  &\epsilon_{\pm 1k_x}=2(m_3-t_2)\sin \frac{k_x}{2}\pm 2(m_2+t_4)\cos\frac{k_x}{2},\qquad {\bm u}_{\pm 1k_x}=\frac{1}{2}(1,\ \mp i,\ e^{i\frac{k_x}{2}},\ \pm ie^{i\frac{k_x}{2}})^{\top},\\
  &\epsilon_{\pm 2k_x}=-2(m_3-t_2)\sin \frac{k_x}{2}\pm 2(m_2+t_4)\cos\frac{k_x}{2},\qquad {\bm u}_{\pm 2k_x}=\frac{1}{2}(1,\ \pm i,\ -e^{i\frac{k_x}{2}},\ \pm ie^{i\frac{k_x}{2}})^{\top}.
\end{align}
By introducing
\begin{gather}
  m_0=2\sqrt{(m_3+t_2)^2+(m_2-t_4)^2},\quad \tan \alpha = \frac{m_2-t_4}{m_3+t_2},
\end{gather}
the eigenenergies can be expressed as
\begin{align}
  &\epsilon_{\pm1k_x}=m_0\sin\left(\frac{k_x}{2}+\alpha\right),\quad \epsilon_{\pm2k_x}=-m_0\sin\left(\frac{k_x}{2}-\alpha\right),
\end{align}
The matrix elements of the current in the basis of these eigenstates, $ \hat{\bm \psi}_{k} = ({\bm u}_{1k}, {\bm u}_{-1k}, {\bm u}_{2k}, {\bm u}_{-2k})\hat{\bm \gamma}_{k}$, are given by
\begin{align}
  j_{k,q}^{\psi}&=e^{-\ii \frac{q}{2}}\cos\left(\frac{q}{4}\right)\sin\left(\frac{2k+q}{4}\right)
  \begin{pmatrix}
    0 & t_1-\ii t_3 & 0 & 0\\
    t_1+\ii t_3 & 0 & 0 & 0\\
    0 & 0 & 0 & -t_1-\ii t_3\\
    0 & 0 & -t_1+\ii t_3 & 0
  \end{pmatrix}\notag\\
  &\quad +e^{-\ii \frac{q}{2}}\sin\left(\frac{q}{4}\right)\cos\left(\frac{2k+q}{4}\right)
  \begin{pmatrix}
    0 & 0 & -t_1+\ii t_3 & 0\\
    0 & 0 & 0 & -t_1-\ii t_3\\
    t_1+\ii t_3 & 0 & 0 & 0\\
    0 & t_1-\ii t_3 & 0 & 0
  \end{pmatrix}.
\end{align}
Therefore, momentum dependent optical conductivity derived from the effective edge theory is
\begin{align}
    \label{eq:sigma_p11a}
  &{\rm Re}[\sigma^{\rm edge}(\omega,q)]\notag\\
  &=\frac{1}{4\omega}\sum_{n,m}\int_{-\pi}^{\pi} dk\{f(\epsilon_{nk_x})-f(\epsilon_{mk_x+q})\}\left|[j_{k_x,q}^{\psi}]_{nm}\right|^2\delta(\omega-\epsilon_{mk_x+q}+\epsilon_{nk_x})\notag\\
  &=\frac{t_1^2+t_3^2}{4\omega m_0}\cos^2\left(\frac{q}{4}\right)\notag\\
  &\quad \times\left[
    \left\{f\left(E_{+}^{1}(q,-\omega)\right)-f\left(E_{+}^{1}(q,\omega)\right)+f\left(E_{-}^{1}(q,-\omega)\right)-f\left(E_{-}^{1}(q,\omega)\right)\right\}\frac{1}{\left|\sin\left(\alpha+\frac{q}{4}\right)\right|}\sqrt{1-\left(\frac{\omega}{2m_0 \sin\left(\alpha+\frac{q}{4}\right)}\right)^2}\right.\notag\\
    &\quad\left.+\left\{f\left(E_{-}^{1}(-q,-\omega)\right)-f\left(E_{-}^{1}(-q,\omega)\right)+f\left(E_{+}^{1}(-q,-\omega)\right)-f\left(E_{+}^{1}(-q,\omega)\right)\right\}\frac{1}{\left|\sin\left(\alpha-\frac{q}{4}\right)\right|}\sqrt{1-\left(\frac{\omega}{2m_0 \sin\left(\alpha-\frac{q}{4}\right)}\right)^2}\right]\notag\\
  &\quad+\frac{t_1^2+t_3^2}{4\omega m_0}\sin^2\left(\frac{q}{4}\right)\notag\\
  &\quad \times\left[
    \left\{f\left(E_{+}^{2}(q,-\omega)\right)-f\left(E_{+}^{2}(q,\omega)\right)+f\left(E_{-}^{2}(q,-\omega)\right)-f\left(E_{-}^{2}(q,\omega)\right)\right\}\frac{1}{\left|\cos\left(\alpha+\frac{q}{4}\right)\right|}\sqrt{1-\left(\frac{\omega}{2m_0 \cos\left(\alpha+\frac{q}{4}\right)}\right)^2}\right.\notag\\
    &\quad\left.+\left\{f\left(E_{+}^{2}(-q,-\omega)\right)-f\left(E_{+}^{2}(-q,\omega)\right)+f\left(E_{-}^{2}(-q,-\omega)\right)-f\left(E_{-}^{2}(-q,\omega)\right)\right\}\frac{1}{\left|\cos\left(\alpha-\frac{q}{4}\right)\right|}\sqrt{1-\left(\frac{\omega}{2m_0 \cos\left(\alpha-\frac{q}{4}\right)}\right)^2}\right].
\end{align}
For simplification, we introduce
\begin{align}
    E^1_{\pm}(q,\omega)=\pm m_0 \cos\left(\alpha+\frac{q}{4}\right)\sqrt{1-\left(\frac{\omega}{2m_0\sin\left(\alpha+\frac{q}{4}\right)}\right)^2}+\frac{\omega}{2},\\
    E^2_{\pm}(q,\omega)=\pm m_0 \sin\left(\alpha+\frac{q}{4}\right)\sqrt{1-\left(\frac{\omega}{2m_0\cos\left(\alpha+\frac{q}{4}\right)}\right)^2}+\frac{\omega}{2}.
\end{align}
Furthermore, setting $q=0$, the optical conductivity is given by
\begin{align}
  &{\rm Re}[\sigma^{\rm edge}(\omega,q=0)]\notag\\
  &=\frac{t_1^2+t_3^2}{2\omega m_0}\frac{1}{\left|\sin\alpha\right|}\sqrt{1-\left(\frac{\omega}{2m_0 \sin\alpha}\right)^2}\notag\\
  &\quad \times\left[
    \left\{f\left(m_0 \cos\alpha\sqrt{1-\left(\frac{\omega}{2m_0\sin\alpha}\right)^2}-\frac{\omega}{2}\right)-f\left(m_0 \cos\alpha\sqrt{1-\left(\frac{\omega}{2m_0 \sin\alpha}\right)^2}+\frac{\omega}{2}\right)\right.\right.\notag\\
    &\qquad \quad \left.\left. +f\left(-m_0 \cos\alpha\sqrt{1-\left(\frac{\omega}{2m_0\sin\alpha}\right)^2}-\frac{\omega}{2}\right)-f\left(-m_0 \cos\alpha\sqrt{1-\left(\frac{\omega}{2m_0 \sin\alpha}\right)^2}+\frac{\omega}{2}\right)\right\}
  \right].
\end{align}
These results correspond to Fig.~8(b,c).

\section{current operator and ${\mathcal C}$ symmetry}
\label{sec:current_C_symmetry}
To consider the relationship between symmetry and optical excitation~[44], we examine the transformation of the current operator under the ${\mathcal C}$ symmetry.
The current operator $\hat{j}^i_{\bm k}$ for the spatially uniform component is given by
\begin{align}
  \hat{j}^i_{\bm k}=\sum_{\bm R}\hat{\tilde{j}}^i_{\bm R}=\left.-\frac{\delta \hat{H}({\bm A})}{\delta A^i}\right|_{{\bm A}={\bm 0}}.
\end{align}
Here, unlike in Eq.~(10), the derivative is taken with respect to a spatially uniform gauge field.
To analyze the Hamiltonian under a uniform gauge field, it is useful to consider a non-periodic basis that takes into account the positions of orbitals within the unit cell.
If we denote by ${\bm r}$ a matrix that lists the positions of orbitals within the unit cell along its diagonal, the non-periodic basis can be expressed as
\begin{align}
  \hat{\bm \Psi}^{\rm np}_{\bm k}=e^{-\ii {\bm k}\cdot{\bm r}}\hat{\bm \Psi}^{}_{\bm k}.
\end{align}
The Hamiltonian in this basis is then written as
\begin{align}
  \hat{H}&=\frac{1}{2}\sum_{\bm k}\hat{\bm \Psi}^{{\rm np}\dagger}_{\bm k}H_{\bm k}^{\rm np}\hat{\bm \Psi}^{\rm np}_{\bm k},\\
  H_{\bm k}^{\rm np}&=e^{-\ii {\bm k}\cdot{\bm r}}H_{\bm k}e^{\ii{\bm k}\cdot{\bm r}}=
    \begin{pmatrix}
      h_{\bm k}^{\rm np} & \Delta_{\bm k}^{\rm np}\\
      \Delta_{\bm k}^{{\rm np}\dagger} & -h_{-{\bm k}}^{{\rm np}\top}
    \end{pmatrix}.
\end{align}
By introducing a uniform gauge field, the Hamiltonian becomes
\begin{align}
  \hat{H}({\bm A})&=\frac{1}{2}\sum_{\bm k}\hat{\bm \Psi}^{{\rm np}\dagger}_{\bm k}
  \begin{pmatrix}
    h_{\bm k}^{\rm np}({\bm A}) & \Delta_{\bm k}^{\rm np}\\
    \Delta_{\bm k}^{{\rm np}\dagger} & -[h_{-{\bm k}}^{{\rm np}}({\bm A})]^{\top}
  \end{pmatrix}
  \hat{\bm \Psi}^{\rm np}_{\bm k}\notag\\
  &=\frac{1}{2}\sum_{\bm k}\hat{\bm \Psi}^{{\rm np}\dagger}_{\bm k}
  \begin{pmatrix}
    h_{{\bm k}+{\bm A}}^{\rm np} & \Delta_{\bm k}^{\rm np}\\
    \Delta_{\bm k}^{{\rm np}\dagger} & -h_{-{\bm k}+{\bm A}}^{{\rm np}\top}
  \end{pmatrix}
  \hat{\bm \Psi}^{\rm np}_{\bm k}.
\end{align}
Thus, the current operator is given by
\begin{align}
  \hat{j}^i_{\bm k}&=\frac{1}{2}\sum_{\bm k}\hat{\bm \Psi}^{{\rm np}\dagger}_{\bm k}
  \begin{pmatrix}
    \partial_{k^i}h_{{\bm k}}^{\rm np} & 0\\
    0 & \partial_{k^i}h_{-{\bm k}}^{{\rm np}\top}
  \end{pmatrix}
  \hat{\bm \Psi}^{\rm np}_{\bm k}\notag\\
  &=\frac{1}{2}\sum_{\bm k}\hat{\bm \Psi}^{\dagger}_{\bm k}
  e^{\ii {\bm k}\cdot{\bm r}}
  \frac{1}{2}(\tau_z\partial_{k^i}H_{\bm k}^{\rm np}+\partial_{k^i}H_{\bm k}^{\rm np}\tau_z)
  e^{-\ii {\bm k}\cdot{\bm r}}\hat{\bm \Psi}^{}_{\bm k}.
\end{align}
Therefore, the matrix element $j^i_{\bm k}$ in the $\hat{\bm \Psi}^{}_{\bm k}$ basis is
\begin{align}
  j^i_{\bm k}=
  \frac{1}{2}e^{\ii {\bm k}\cdot{\bm r}}(\tau_z\partial_{k^i}H_{\bm k}^{\rm np}+\partial_{k^i}H_{\bm k}^{\rm np}\tau_z)
  e^{-\ii {\bm k}\cdot{\bm r}}.
\end{align}

Next, we consider how the current operator transforms under the symmetry ${\mathcal C}$.
Since ${\mathcal C}$ is an anti-unitary symmetry that does not change ${\bm k}$, its representation is
\begin{align}
  U_{\bm k}^{\rm np}({\mathcal C})=e^{-\ii {\bm k}\cdot{\bm r}}U_{\bm k}({\mathcal C})e^{-\ii {\bm k}\cdot{\bm r}}
\end{align}
and it satisfies
\begin{align}
  U_{\bm k}^{\rm np}({\mathcal C})H^{{\rm np}\top}_{\bm k}=-H^{\rm np}_{\bm k}U_{\bm k}^{\rm np}({\mathcal C}).
\end{align}
Generally, the representation of a symmetry in a non-periodic basis can separate the momentum-dependent part, and $U_{\bm k}^{\rm np}({\mathcal C})$ can be expressed as
\begin{align}
  U_{\bm k}^{\rm np}({\mathcal C})=e^{-\ii{\bm k}\cdot{\bm t}_{\mathcal C}}U({\bm C}),
\end{align}
where ${\bm t}_{\mathcal C}$ represents the translational part of ${\mathcal C}$.
Now, considering the transformation of the current operator under ${\mathcal C}$, we have
\begin{align}
  &\quad U_{\bm k}({\mathcal C})(j^i_{\bm k})^*U^{\dagger}_{\bm k}({\mathcal C})\notag\\
  &=e^{\ii{\bm k}\cdot{\bm r}}U^{\rm np}_{\bm k}({\mathcal C})e^{\ii{\bm k}\cdot{\bm r}}\frac{1}{2}e^{-\ii{\bm k}\cdot{\bm r}}(\tau_z\partial_{k^i}H_{\bm k}^{{\rm np}*}+\partial_{k^i}H_{\bm k}^{{\rm np}*}\tau_z)
  e^{\ii{\bm k}\cdot{\bm r}}e^{-\ii{\bm k}\cdot{\bm r}}U^{{\rm np}\dagger}_{\bm k}({\mathcal C})e^{-\ii{\bm k}\cdot{\bm r}}\notag\\
  &=\frac{1}{2}e^{\ii{\bm k}\cdot{\bm r}}U^{\rm np}_{\bm k}({\mathcal C})(\tau_z\partial_{k^i}H_{\bm k}^{{\rm np}*}+\partial_{k^i}H_{\bm k}^{{\rm np}*}\tau_z)
  U^{{\rm np}\dagger}_{\bm k}({\mathcal C})e^{-\ii{\bm k}\cdot{\bm r}}\notag\\
  &=-\frac{1}{2}e^{\ii{\bm k}\cdot{\bm r}}\{\tau_z U^{\rm np}_{\bm k}({\mathcal C})(\partial_{k^i}H_{\bm k}^{{\rm np}*})U^{{\rm np}\dagger}_{\bm k}({\mathcal C})+U^{\rm np}_{\bm k}({\mathcal C})(\partial_{k^i}H_{\bm k}^{{\rm np}*})U^{{\rm np}\dagger}_{\bm k}({\mathcal C})\tau_z\}
  e^{-\ii{\bm k}\cdot{\bm r}}.
\end{align}
Considering that
\begin{align}
  &\quad U^{\rm np}_{\bm k}({\mathcal C})(\partial_{k^i}H_{\bm k}^{{\rm np}*})U^{{\rm np}\dagger}_{\bm k}({\mathcal C})\notag\\
  &=\partial_{k^i}[U^{\rm np}_{\bm k}({\mathcal C})H_{\bm k}^{{\rm np}*}U^{{\rm np}\dagger}_{\bm k}({\mathcal C})]-\partial_{k^i}[U^{\rm np}_{\bm k}({\mathcal C})]H_{\bm k}^{{\rm np}*}U^{{\rm np}\dagger}_{\bm k}({\mathcal C})-U^{\rm np}_{\bm k}({\mathcal C})H_{\bm k}^{{\rm np}*}\partial_{k^i}[U^{{\rm np}\dagger}_{\bm k}({\mathcal C})]\notag\\
  &=\partial_{k^i}[-H_{\bm k}^{{\rm np}}]-(-\ii t_{\mathcal C}^i)U^{\rm np}_{\bm k}({\mathcal C})H_{\bm k}^{{\rm np}*}U^{{\rm np}\dagger}_{\bm k}({\mathcal C})-(+\ii t_{\mathcal C}^i)U^{\rm np}_{\bm k}({\mathcal C})H_{\bm k}^{{\rm np}*}U^{{\rm np}\dagger}_{\bm k}({\mathcal C})\notag\\
  &=-\partial_{k^i}H_{\bm k}^{{\rm np}},
\end{align}
we find that
\begin{align}
  &\quad U_{\bm k}({\mathcal C})(j^i_{\bm k})^*U_{\bm k}^{\dagger}({\mathcal C})=j^i_{\bm k}.
\end{align}

\end{document}